\documentclass{SciPost}
\hypersetup{
    colorlinks,
    linkcolor={red!50!black},
    citecolor={blue!50!black},
    urlcolor={blue!80!black}
}

\usepackage[bitstream-charter]{mathdesign}
\usepackage[section]{placeins}

\DeclareSymbolFont{usualmathcal}{OMS}{cmsy}{m}{n}
\DeclareSymbolFontAlphabet{\mathcal}{usualmathcal}

\fancypagestyle{SPstyle}{
\fancyhf{}
\lhead{\colorbox{scipostblue}{\bf \color{white} ~SciPost Physics }}
\rhead{{\bf \color{scipostdeepblue} ~Submission }}

\fancyfoot[C]{\textbf{\thepage}}
}

\begin{document}

\pagestyle{SPstyle}

\begin{center}{\Large \textbf{\color{scipostdeepblue}{Finite-Temperature Transconductance of Multiterminal Josephson Junctions with Topological Andreev Spectra
}}}\end{center}

\begin{center}\textbf{
Gleb S. Seleznev\textsuperscript{1$\star$},
David C. Ohnmacht\textsuperscript{1,$\dagger$} and
Wolfgang Belzig\textsuperscript{1}
}\end{center}


\begin{center}
\textbf{1} Fachbereich Physik, University of Konstanz, D-78457 Konstanz, Germany
\par\vspace{\baselineskip}
\ensuremath{\star} \href{mailto:gleb.seleznev@uni-konstanz.de}{\small gleb.seleznev@uni-konstanz.de}
\end{center}

\begingroup
\renewcommand{\thefootnote}{\ensuremath{\dagger}}
\footnotetext[2]{Current affiliation: QuTech and Kavli Institute of Nanoscience, Delft University of Technology, 2628 CJ Delft, The Netherlands.}
\endgroup

\section*{\color{scipostdeepblue}{Abstract}}
\textbf{\boldmath{%
Multiterminal Josephson junctions (MTJJs) provide a controllable platform for engineering topological Andreev spectra. At zero temperature, Andreev bound states (ABSs) with nonzero Chern numbers give rise to quantized transconductance between superconducting terminals. We derive the finite-temperature transconductance of MTJJs using linear-response theory. We first demonstrate the destructive role of temperature: changes in the thermal occupations of the ABSs destroy exact transconductance quantization. We then show that finite temperature can also play a constructive role by making the transconductance sensitive to finite-energy topological crossings hidden from the zero-temperature response. Our results establish the thermal limitations of transconductance quantization and identify finite-temperature transport as a probe of otherwise inaccessible topology in Andreev spectra.}}

\vspace{\baselineskip}

\noindent\textcolor{white!90!black}{%
\fbox{\parbox{0.975\linewidth}{%
\textcolor{white!40!black}{\begin{tabular}{lr}%
  \begin{minipage}{0.6\textwidth}%
    {\small Copyright attribution to authors. \newline
    This work is a submission to SciPost Physics. \newline
    License information to appear upon publication. \newline
    Publication information to appear upon publication.}
  \end{minipage} & \begin{minipage}{0.4\textwidth}
    {\small Received Date \newline Accepted Date \newline Published Date}%
  \end{minipage}
\end{tabular}}
}}
}

\vspace{10pt}
\noindent\rule{\textwidth}{1pt}
\tableofcontents
\noindent\rule{\textwidth}{1pt}
\vspace{10pt}

\section{Introduction}
\label{sec:intro}
Topological band theory provides a framework for phases whose transport can be governed by integer topological invariants~\cite{BansilRevModPhys2016}. Representative examples include topological insulators~\cite{HasanRevModPhys2010}, Weyl semimetals~\cite{Fesler2017}, and topological superconductors~\cite{Sato2017}. The paradigmatic quantized response is the integer quantum Hall effect, for which $\sigma_{xy}=(e^2/h)\nu$ ~\cite{Klitzing1980,Thouless1982}. Here, $\nu$ is an integer Chern number, $e$ is the electron charge, and $h$ is Planck's constant.

Synthetic topological systems realize analogous band structures in controllable parameter spaces rather than exclusively in real space~\cite{Ozawa2019,Arguello2024}. Examples include Floquet systems~\cite{Kitagaws2010,Rudner2020}, photonic platforms~\cite{Khanikaev2013,Lu2014}, and superconducting circuits~\cite{Fatemi2021,Peyruchat2021,Herrig2022}. Their tunability also enables effective topological physics in dimensions that are difficult to access in ordinary materials~\cite{Kraus2013,Price2015,Lohse2018,Zilberberg2018}.

Multiterminal Josephson junctions (MTJJs) are another tunable platform for exploring unconventional superconducting transport and phase-controlled Andreev physics \cite{Cuevas2007, Houzet2010, Freyn2011, Pfeffer2014, Cohen2018,  Draelos2019, Pillet2019,  Graziano2020, Pankratove2020, Arnault2021, Huang2022, Graziano2022, Matsuo2022, Gupta2023, Haxell2023, Coraiola2023, Matsuo2023P, Coraiola2024diode, Prosko2024,  Virtanen2024, Kahn2025, Annabi2026} and, in particular, synthetic topology ~\cite{Yokoyama2015, Riwar2016,Eriksson2017,Meyer2017, Xie2017, Xie2018,  Xie2019, Repin2019, Klees2020,Klees2021, Chen2021, Meyer2021, Weisbrich2021CN ,Xie2022, Barakov2023, Gavensky2023,Teshler2023, Antonelli2025, Ohnmacht2025NH, Ohnmacht2025RM, Jung2025, Chakraborty2026, Zalom2026}. In an $N$-terminal junction, the $N-1$ independent superconducting phase differences act as synthetic quasimomenta, while the Andreev bound states (ABSs) form an effective band structure in this phase space. For $N\geq4$, the phase space is at least three-dimensional and can host Weyl points (WPs), i.e., conical degeneracies carrying integer topological charge~\cite{Yokoyama2015,Riwar2016,Xie2018}. These WPs can result in a nonzero ground-state Chern number or redistribute Chern numbers among individual ABSs.

In the adiabatic transport regime, MTJJ topology manifests itself in the dc transconductance. Incommensurate voltages applied to two superconducting (SC) leads make the corresponding SC phases densely cover a two-dimensional phase Brillouin zone (BZ), i.e., a torus. The BZ-averaged transconductance is then quantized as ${G}^{ij}=(4e^2/h)C^{ij}$, where $C^{ij}$ is the ground-state Chern number. This is the synthetic-phase analogue of the quantized Hall response \cite{Riwar2016}.

Experimental observation of this quantization is challenging. Finite bias can drive nonadiabatic ABS transitions and restrict the adiabatic voltage window~\cite{Eriksson2017}, while level broadening can smear the response~\cite{Chen2021}. The measurement is also intrinsically nonlocal because it correlates a voltage on one terminal with the current in another. These limitations have motivated complementary probes, including tunneling~\cite{Antonelli2025,Jung2025} and microwave spectroscopy~\cite{Klees2020,Klees2021}.

Quantization can nevertheless be robust in special regimes, for example with Majorana-hosting SC terminals~\cite{Meyer2021} or when the scattering matrix of the normal region lies near a chiral point~\cite{Chakraborty2026}. For time-reversal-symmetric junctions, continuum states above the SC gap $\Delta$ can also preserve a quantized contribution~\cite{Repin2019}.

A further open question concerns the effect of finite temperature on transconductance quantization. At finite temperature, ABS occupations are no longer step functions, so different states acquire unequal thermal weights. How this thermal reweighting modifies the quantized response has not been systematically investigated. In addition, it remains unclear whether transconductance can probe the topology of individual ABSs rather than only that of the ground state. This question is particularly relevant when individual ABSs carry nonzero Chern numbers whose sum vanishes, so that the zero-temperature transconductance contains no signature of their individual topology.

Here we address both questions in a four-terminal MTJJ within the framework of effective quantum dot (QD) models \cite{Klees2020, Teshler2023}. First, we determine how thermal occupation modifies the quantized transconductance and show that exact quantization is lost once the temperature becomes comparable to the relevant ABS energy scale. Second, we show that finite-temperature transconductance can reveal individual-ABS topology that is absent from the zero-temperature response. Thus, the same thermal reweighting that limits quantization can also expose otherwise hidden structure of the Andreev spectrum.

The paper is organized as follows. In Sec.~\ref{sec: model} we discuss the synthetic topology of MTJJs. In Sec.~\ref{sec: effective QD models} we introduce the three effective QD models used throughout the paper and summarize their topological properties. In Sec.~\ref{sec: linear response} we derive the finite-temperature transconductance within linear-response theory. In Sec.~\ref{sec: violation of quantization} we demonstrate the thermal breakdown of the transconductance quantization. In Sec.~\ref{sec: Detection of FEWPs} we investigate the finite-temperature signatures of individual-ABS topology. In Sec.~\ref{sec: Coexistence of ZEWPs and FEWPs} we discuss the coexistence of ground-state and individual-ABS topology.  In Sec.~\ref{sec:discussion} we discuss the experimental implications. In  Sec.~\ref{sec:conclusions} we present our conclusions. Technical derivations are collected in the Appendices.

\section{Topology of MTJJs}\label{sec: model}

We first summarize the synthetic topology of MTJJs. This section concerns the zero-temperature limit, $T=0$; the finite-temperature response is introduced in Sec.~\ref{sec: linear response}.

\subsection{Synthetic topology}

\begin{figure*}[tbp]
\centering
\includegraphics[width=0.5\columnwidth]{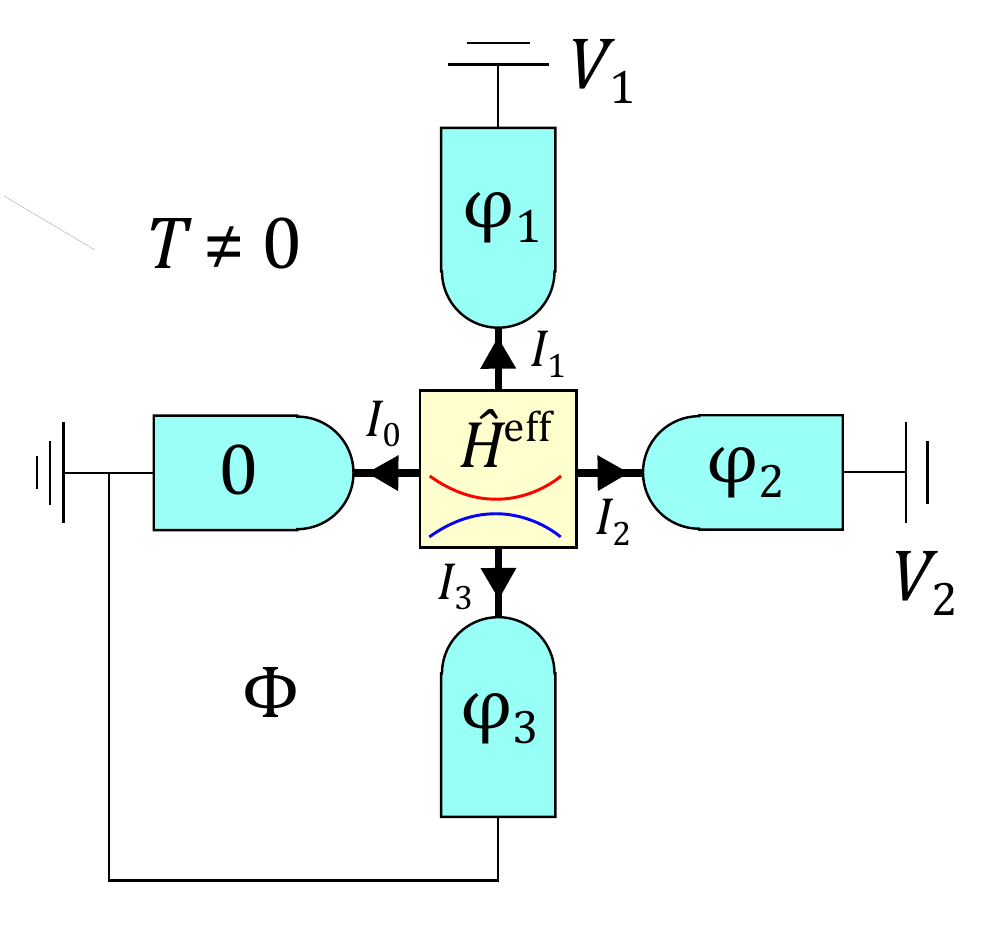}
\caption{Schematic four-terminal MTJJ. Four $s$-wave SC leads with phases $\varphi_\alpha$ ($\alpha=0,\ldots,3$) are coupled through a normal region hosting the ABSs described by $\hat H^{\mathrm{eff}}$. We choose the gauge $\varphi_0=0$. Voltages $V_1$ and $V_2$ drive $\varphi_1$ and $\varphi_2$, while $\varphi_3$ is controlled by the flux $\Phi$ through the loop connecting leads $0$ and $3$. The current $I_\alpha$ is measured in lead $\alpha$, and the junction is at temperature $T\neq0$.}
\label{fig: Scheme}
\end{figure*}

We consider $N$ SC terminals coupled through a normal region; see Fig.~\ref{fig: Scheme} for an example of four-terminal geometry. SC terminals $\alpha=0,\ldots,N-1$ are $s$-wave superconductors with gaps $\Delta_\alpha$ and phases $\varphi_\alpha$. Gauge invariance leaves $N-1$ independent SC phase differences, and we set $\varphi_0=0$. We assume $M$ relevant ABSs, which are described by a $2M\times2M$ Bogoliubov--de Gennes (BdG) Hamiltonian $\hat H^{\mathrm{eff}}$. Particle-hole symmetry gives $M$ negative-energy states and $M$ positive-energy partners. Energies are measured relative to the common chemical potential $E_F$. 

The matrix elements of $\hat H^{\mathrm{eff}}$ depend on the SC
phases $\varphi_\alpha$. These phases define a synthetic parameter space, in which
the ABS spectrum can acquire nontrivial topology. For the
$k$th ABS eigenstate $|k\rangle$ of $\hat H^{\mathrm{eff}}$, with energy
$E_k$, we define the Berry curvature in the $(\varphi_i,\varphi_j)$ phase
plane as
\begin{equation}
    B_k^{ij}
    =
    2\,\mathrm{Im}
    \left[
        \left(\partial_{\varphi_i}\langle k|\right)
        \partial_{\varphi_j}|k\rangle
    \right],
    \label{eq:Berry_curvature_ABS}
\end{equation}
and the corresponding Chern number as
\begin{equation}
    C_k^{ij}
    =
    \frac{1}{2\pi}
    \int_0^{2\pi} d\varphi_i
    \int_0^{2\pi} d\varphi_j\,
    B_k^{ij}.
    \label{eq:Chern_number_ABS}
\end{equation}
The derivatives and integrals are taken over the two SC phases spanning the chosen two-dimensional synthetic BZ.

At $T=0$, the ground-state Berry curvature and Chern number are obtained by summing over the occupied negative-energy states:
\begin{equation}\label{eq: BC and CN}
   B^{ij}
   =
   \sum_{E_k<0} B_k^{ij},
   \qquad
   C^{ij}
   =
   \sum_{E_k<0} C_k^{ij},
\end{equation}
A nonzero $C^{ij}_k$ identifies a topologically nontrivial $k$th ABS, whereas $C^{ij}\neq0$ identifies a topologically nontrivial ground state. Importantly, individual ABSs can carry nonzero Chern numbers whose sum cancels, yielding a topologically trivial ground state with $C^{ij}=0$.

We now focus on a four-terminal MTJJ ($N=4$). Topology is evaluated on the $(\varphi_1,\varphi_2)$ phase torus, with $\varphi_3$ as a control parameter, so we study $C_k^{12}(\varphi_3)$ and $C^{12}(\varphi_3)$. We call a parameter regime topological when at least one ABS has $C_k^{12}\neq0$ over a finite interval of $\varphi_3$. We distinguish a \emph{ground-state topological} regime, $C^{12}\neq0$, from an \emph{individual-ABS topological} regime, $C^{12}=0$ and $C_k^{12}\neq0$ for at least one band.

\subsection{Different types of WPs}\label{sec: different types of WPs}

\begin{figure*}[tbp]
\centering
\includegraphics[width=0.4\columnwidth]{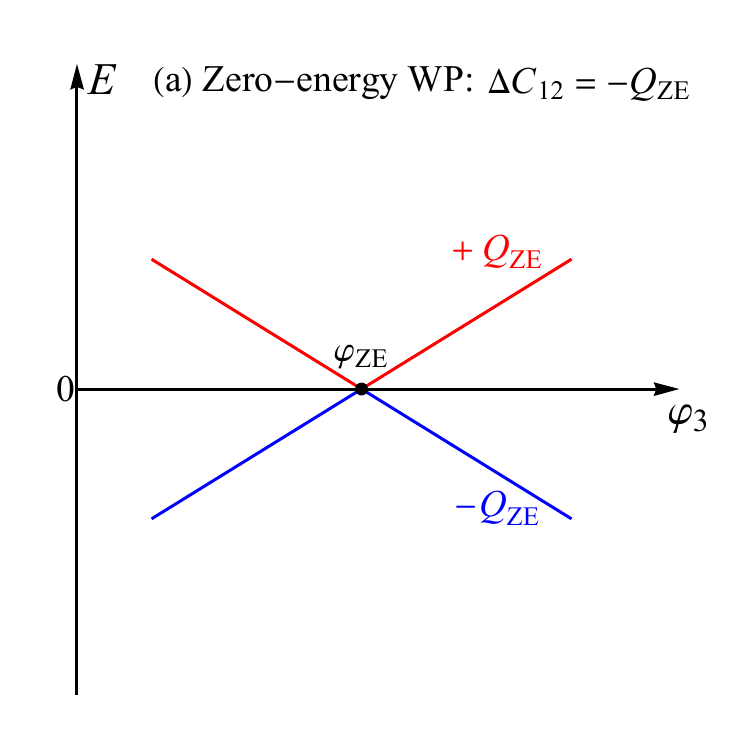} 
\hspace{1 cm}
\includegraphics[width=0.4\columnwidth]{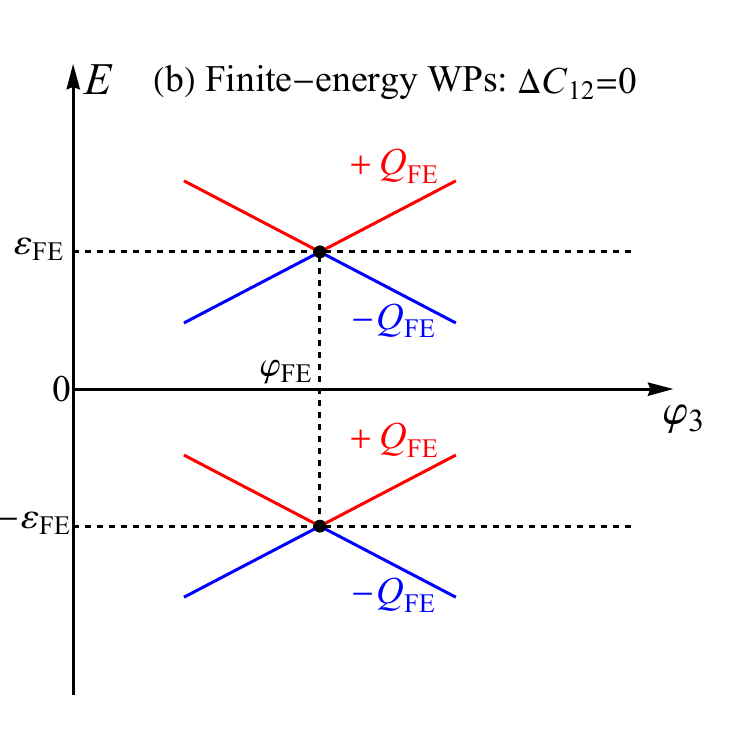}
\caption{Schematic ABS spectra versus the control phase $\varphi_3$ at fixed $\varphi_1$ and $\varphi_2$. At $T=0$, states with $E<0$ ($E>0$) are occupied (unoccupied). (a) At the zero-energy WP $\varphi_3=\varphi_{\mathrm{ZE}}$, the negative- and positive-energy branches change their Chern numbers by $-Q_{\mathrm{ZE}}$ and $+Q_{\mathrm{ZE}}$, giving $\Delta C^{12}=-Q_{\mathrm{ZE}}$. (b) Particle-hole-related finite-energy WPs occur at $E=\pm\varepsilon_{\mathrm{FE}}$ and $\varphi_3=\varphi_{\mathrm{FE}}$. At each crossing, the lower and upper branches change their Chern numbers by $-Q_{\mathrm{FE}}$ and $+Q_{\mathrm{FE}}$, while the ground-state Chern number remains unchanged, $\Delta C^{12}=0$.}
\label{fig: Schematic WPs}
\end{figure*}

The synthetic topology of MTJJs is governed by WPs in the ABS spectrum~\cite{Riwar2016}. We distinguish two types. Zero-energy WPs occur when a positive-energy ABS meets its particle-hole partner at $E=0$; see Fig.~\ref{fig: Schematic WPs}(a). Near a zero-energy WP at $\boldsymbol{\varphi}_{\mathrm{ZE}}$, the effective Hamiltonian is 
\begin{equation}
\label{eq:ZEWP}
    \hat{H}_{\mathrm{ZE}}
    =
    \mathbf{d}_{\mathrm{ZE}}
    \cdot\boldsymbol{\hat{\tau}},
    \qquad
    \mathbf{d}_{\mathrm{ZE}}
    =
    \hat{M}_{\mathrm{ZE}}\delta\boldsymbol{\varphi},
\end{equation}
where $\delta\boldsymbol{\varphi}=\boldsymbol{\varphi}-\boldsymbol{\varphi}_{\mathrm{ZE}}$, $\boldsymbol{\hat{\tau}}=(\hat{\tau}_x,\hat{\tau}_y,\hat{\tau}_z)$ is a vector of Pauli matrices, and $\hat M_{\mathrm{ZE}}$ is a real $3\times3$ velocity matrix. A zero-energy WP has topological charge
\begin{equation}\label{eq: topological charge}
Q_{\mathrm{ZE}}= \operatorname{sgn}\det \hat{M}_{\mathrm{ZE}}. 
\end{equation}

Crossing the zero-energy WP as $\varphi_3$ changes the Chern numbers of the negative- and positive-energy branches by $-Q_{\mathrm{ZE}}$ and $+Q_{\mathrm{ZE}}$, respectively. At $T=0$, only the former is occupied, so the ground-state Chern number changes by $-Q_{\mathrm{ZE}}$.

Finite-energy WPs occur when two ABSs within the same $T=0$ occupation sector cross; see Fig.~\ref{fig: Schematic WPs}(b). Particle-hole symmetry generates a partner crossing at the opposite energy. In a particle-hole-related basis, the four relevant BdG states are described by
\begin{equation}
\label{eq:FEWP}
    \hat{H}_{\mathrm{FE}}
    =
    \begin{pmatrix}
        \hat{h}_{\mathrm{FE}} & 0 \\
        0 & -\hat{h}_{\mathrm{FE}}^{*}
    \end{pmatrix}, \quad     \hat{h}_{\mathrm{FE}}
    =
    E_{\mathrm{FE}}\hat{\tau}_0
    +
    \mathbf{d}_{\mathrm{FE}}
    \cdot\boldsymbol{\hat{\tau}}
\end{equation}
where 
\begin{gather}
\label{eq:FEWP_block}
    E_{\mathrm{FE}}
    =
    \varepsilon_{\mathrm{FE}}
    +
    \mathbf{v}_{\mathrm{FE}}\cdot
    \delta\boldsymbol{\varphi}, \quad \mathbf{d}_{\mathrm{FE}}
    =
    \hat{M}_{\mathrm{FE}}\delta\boldsymbol{\varphi},
\end{gather}
Here $\varepsilon_{\mathrm{FE}}>0$ is the positive-energy crossing energy, $\mathbf v_{\mathrm{FE}}$ describes the cone tilt, $\hat M_{\mathrm{FE}}$ is a $3\times3$ velocity matrix, and $\delta\boldsymbol{\varphi}=\boldsymbol{\varphi}-\boldsymbol{\varphi}_{\mathrm{FE}}$. The topology is encoded in $\mathbf d_{\mathrm{FE}}$, giving the finite-energy WP charge
\begin{equation}
    Q_{\mathrm{FE}} = \textrm{sgn} \det \hat{M}_{\mathrm{FE}}.
\end{equation}

The particle-hole-related finite-energy WPs at $\pm E_{\mathrm{FE}}$ carry the same charge $Q_{\mathrm{FE}}$. At each crossing, the lower and upper branches change their Chern numbers by $-Q_{\mathrm{FE}}$ and $+Q_{\mathrm{FE}}$. Because both negative-energy branches are occupied at $T=0$, these changes cancel in the ground-state sum. Finite-energy WPs therefore redistribute Chern numbers among individual ABSs without changing the ground-state Chern number.

Thus zero-energy WPs change the ground-state Chern number, whereas finite-energy WPs redistribute Chern numbers among individual ABSs while leaving the ground-state Chern number unchanged. Importantly, finite-energy WPs can result in  nontrivial individual-ABS topology even when the ground state is topologically trivial.

\section{Effective QD models}\label{sec: effective QD models}

We model the ABS spectra using effective QD models~\cite{Klees2020, Teshler2023}, with $M$ single-level QDs tunnel coupled to $N$ SC leads. Integrating out the leads in the atomic limit, $\Delta_\alpha\rightarrow\infty$, yields an effective BdG Hamiltonian $\hat H_D^{\mathrm{eff}}$ for resonant ABSs with $E\ll\Delta_\alpha$; Appendix~\ref{appendix: derivation of low energy Hamiltonian} gives the derivation. We compute the Chern numbers of the ABSs from the effective Hamiltonians using the discretized Fukui algorithm~\cite{Fukui2005}.

\begin{figure*}[tbp]
\centering
\includegraphics[width=1\columnwidth]{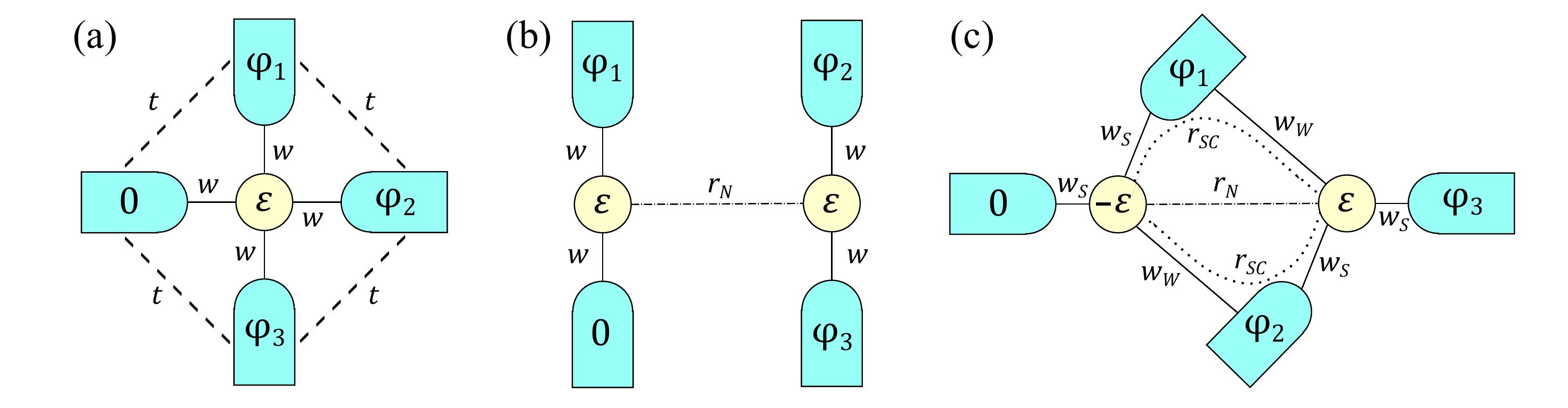}
\caption{Four-terminal QD models: (a) symmetric single-QD, (b) symmetric normal-coupled double-QD, and (c) antisymmetric SC-coupled double-QD. Each QD represents one fermionic level. Panels (a) and (b) use equal on-site energies $\varepsilon$ and dot--lead couplings $w$; panel (c) uses opposite dot energies and strong/weak couplings $w_S$ and $w_W<w_S$. Both double-QD models contain normal hopping $r_N$; the shared leads $1$ and $2$ in panel (c) additionally generate the nonlocal SC coupling $r_{\mathrm{SC}}$. Neighboring SC leads in panel (a) are directly coupled by $t$.}
\label{fig: effective models}
\end{figure*}

We focus on three four-terminal QD models with distinct topological properties. The symmetric single-QD model ($M=1$) is the minimal realization of nontrivial ground-state topology~\cite{Klees2020}; see Fig.~\ref{fig: effective models}(a). The symmetric normal-coupled double-QD model ($M=2$) provides a minimal multiband realization of ground-state topology~\cite{Teshler2023}; see Fig.~\ref{fig: effective models}(b). Finally, the antisymmetric SC-coupled double-QD model supports finite-energy WPs and hence nontrivial individual-ABS topology; see Fig.~\ref{fig: effective models}(c). Appendix~\ref{appendix: Different QD models} summarizes the corresponding effective Hamiltonians.

\subsection{Symmetric single-QD model}\label{sec: SQD model}

The symmetric single-QD model shown in Fig.~\ref{fig: effective models}(a) is the four-terminal model with one topological resonant ABS ~\cite{Klees2020}. A dot with on-site energy $\varepsilon$ couples equally to the four SC terminals with amplitude $w$, while neighboring terminals are directly coupled by $t$. The effective dot Hamiltonian is
\begin{equation}\label{eq: SQD model}
    \hat H_D^{\mathrm{eff}}=\mathbf d\cdot\hat{\boldsymbol\tau}, 
\end{equation}
with
\begin{equation}
    \mathbf{d} =
    \begin{pmatrix}
        \sum^3_{a=0} \Gamma  \cos \varphi_a \\
        - \sum^3_{a=0} \Gamma \sin \varphi_a \\
        \varepsilon
        - 2  \tau \Gamma\sum^3_{a=0}
        \cos (\varphi_a - \varphi_{a+1})
    \end{pmatrix}.
    \label{eq: symmetric SQD model}
\end{equation}
Here $\Gamma=\pi N_0w^2$, $\tau=\pi N_0t$, and $N_0$ is the normal-state density of states per spin at the Fermi level. The sum in Eq.\ \eqref{eq: symmetric SQD model} assumes that the index $a$ is cyclic, so that $\varphi_4 \equiv \varphi_0$.

\begin{figure*}[t]
\includegraphics[width=0.33\columnwidth]{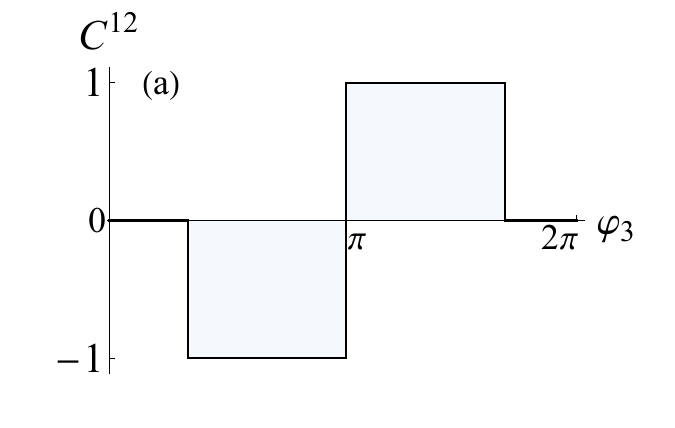}\centering \hfill
\includegraphics[width=0.33\columnwidth]{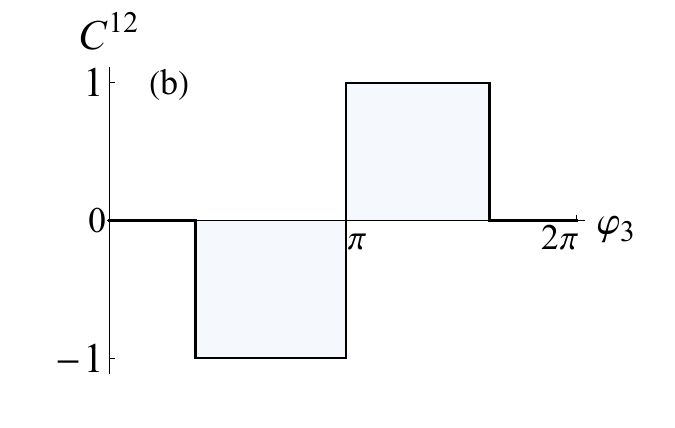}\centering \hfill
\includegraphics[width=0.33\columnwidth]{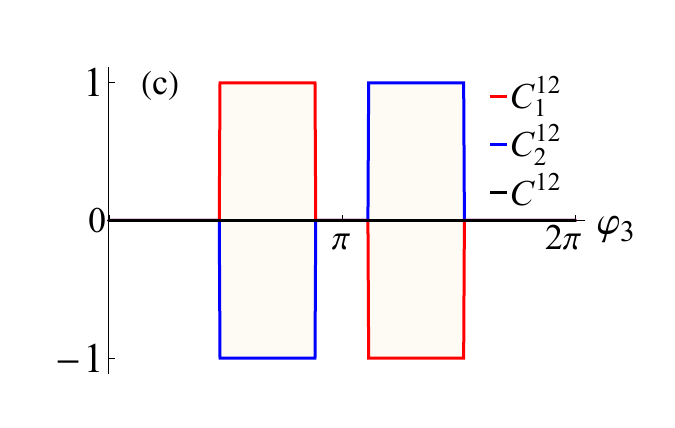}\centering\\
\includegraphics[width=0.33\columnwidth]{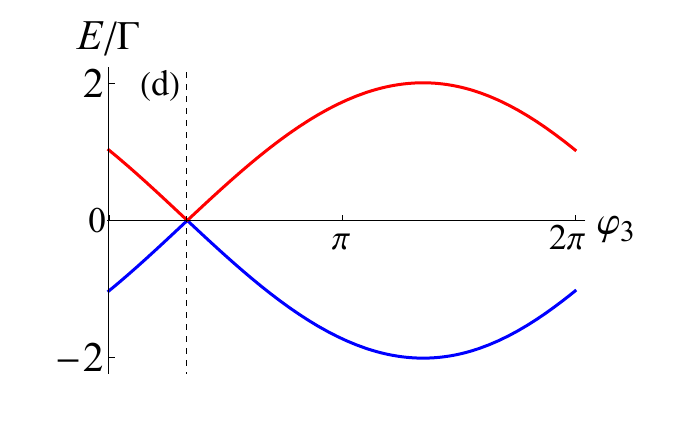}\centering 
\hfill
\includegraphics[width=0.33\columnwidth]{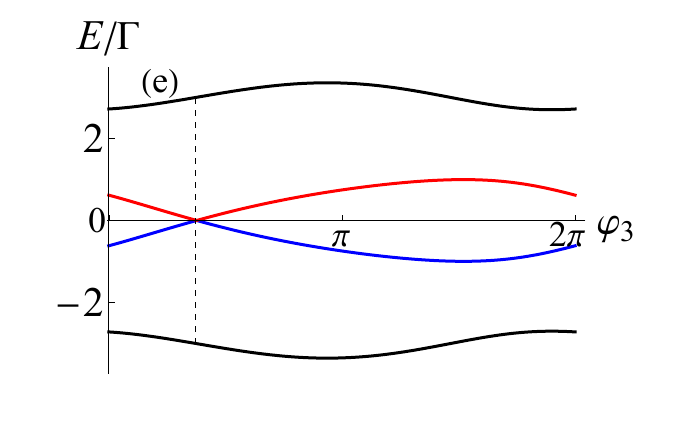}\centering 
\hfill
\includegraphics[width=0.33\columnwidth]{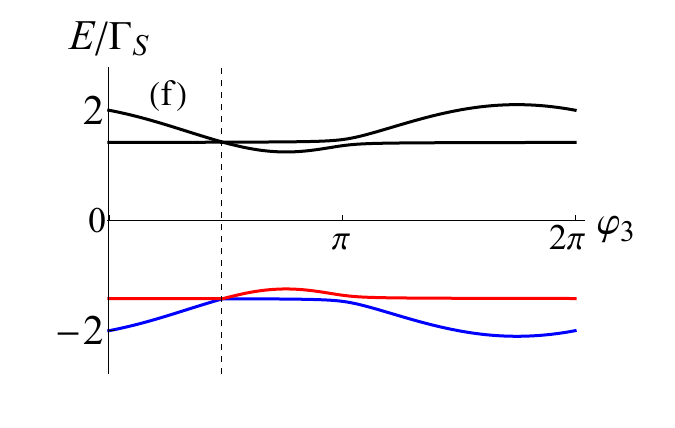}\centering
\caption{Chern numbers (a)--(c) and representative ABS spectra (d)--(f) versus $\varphi_3$. Ground-state and individual-ABS topological regions are shaded light blue and light red, respectively. Red and blue denote the ABS branches whose Chern numbers are shown; additional branches are black. Dashed lines in (d)--(f) mark the first Chern-number jumps in (a)--(c). (a,d) Symmetric single-QD: $\varepsilon/\Gamma=-0.2$, $\tau=0.1$; panel (d), at $\varphi_1=\pi$, $\varphi_2=4\pi/3$, shows a zero-energy WP. (b,e) Symmetric normal-coupled double-QD: $\varepsilon/\Gamma=1$, $r_N/\Gamma=1.5$; panel (e), at $\varphi_1=4.33$, $\varphi_2=\pi$, shows a zero-energy WP. (c,f) Antisymmetric SC-coupled double-QD: $\varepsilon/\Gamma_S=1.2$, $\Gamma_W/\Gamma_S=0.3$, $r_N/\Gamma_S=0.1$. Panel (c) shows $C^{12}$, $C_1^{12}$, and $C_2^{12}$, with $C_1^{12}=-C_2^{12}$ and $C^{12}=0$ in the individual-ABS topological region. Panel (f), at $\varphi_1=2.35$, $\varphi_2=5.5$, shows the occupied-sector finite-energy WP and its particle-hole-related crossing in the unoccupied sector (black).}
\label{fig: Chern numbers and WPs}
\end{figure*}

The ground-state Chern number $C^{12}(\varphi_3)$ and a representative zero-energy WP are shown in Figs.~\ref{fig: Chern numbers and WPs}(a) and \ref{fig: Chern numbers and WPs}(d), respectively. Because the single-QD model has only one positive-energy ABS [red curve in Fig.~\ref{fig: Chern numbers and WPs}(d)], it cannot host finite-energy WPs, whereas zero-energy WPs occur over a broad parameter range; see Appendix~\ref{appendix: ZEWPs in SQD model}.

\subsection{Double-QD model}\label{sec: DQD model}
 To study multiband topology ($M>1$), in particular individual-ABS topology, we next consider double-QD models with two resonant ABS pairs. A general double-QD model consisting of two QDs coupled to SC leads has an effective BdG Hamiltonian of the block form
\begin{equation}
\hat{H}^{\mathrm{eff}}_{D}=
\begin{pmatrix}
    \hat{H}_{1} &  \hat{V}_{12}  \\
    \hat{V}^{\dagger}_{12} & \hat{H}_{2}
\end{pmatrix},
\label{eq: general DQD model}
\end{equation}
where the diagonal and off-diagonal blocks respectively are given by 
\begin{equation}\label{eq: blocks of DQD model}
    \hat{H}_{i} =
    \begin{pmatrix}
        \varepsilon_i & \gamma_{i} \\
        \gamma_i^* & -\varepsilon_i
    \end{pmatrix}, \quad 
    \hat{V}_{12} =
    \begin{pmatrix}
        r_N & r_{\mathrm{SC}} \\
        r_{\mathrm{SC}}^{*} & -r_N
    \end{pmatrix}.
\end{equation}
Here $\varepsilon_i$ and $\gamma_i$ are the on-site energy and induced pairing amplitude on dot $i=1,2$, respectively; $r_N$ is the normal interdot hopping and $r_{\mathrm{SC}}$ is a nonlocal SC interdot coupling generated when the dots share  common SC leads.

We consider two double-QD geometries. In the normal-coupled model [Fig.~\ref{fig: effective models}(b)], the two QDs couple to disjoint pairs of SC leads, so $r_{\mathrm{SC}}=0$ while $r_N\neq0$. In the SC-coupled model [Fig.~\ref{fig: effective models}(c)], the QDs share SC leads, which generates a nonlocal SC coupling $r_{\mathrm{SC}}\neq0$ in addition to $r_N\neq0$. As we will see below, the difference in interdot coupling leads to qualitatively different ABS topology in these two models.

\subsubsection{Symmetric normal-coupled double-QD model}\label{sec: N-DQD model}

We first consider the symmetric normal-coupled double-QD model shown in Fig.~\ref{fig: effective models}(b). It contains two dots with equal on-site energies \cite{Teshler2023}. Dot 1 couples to terminals $0$ and $1$, and dot 2 to terminals $2$ and $3$, all with amplitude $w$. Thus, in Eq.~\eqref{eq: blocks of DQD model}, $\varepsilon_1=\varepsilon_2=\varepsilon$, $\gamma_1=\Gamma(1+e^{i\varphi_1})$, and $\gamma_2=\Gamma(e^{i\varphi_2}+e^{i\varphi_3})$, with $\Gamma=\pi N_0w^2$. The dots are coupled only by the normal hopping $r_N$, so $r_{\mathrm{SC}}=0$.

This model describes two ABSs primarily formed between leads \((0,1)\) and \((2,3)\), respectively, while the normal interdot hopping \(r_N\) introduces a weak hybridization between these two states. Over a finite range of $r_N$, this hybridization produces zero-energy WPs and a nontrivial ground-state Chern number~\cite{Teshler2023}. Figures~\ref{fig: Chern numbers and WPs}(b) and \ref{fig: Chern numbers and WPs}(e) show $C^{12}(\varphi_3)$ and a representative zero-energy WP. The lowest positive- and negative-energy ABSs [red and blue curves in Fig.~\ref{fig: Chern numbers and WPs}(e), respectively] acquire nonzero Chern numbers through zero-energy WPs, whereas finite-energy WPs are absent and the higher-energy pair  [black curves in Fig.~\ref{fig: Chern numbers and WPs}(e)] remains topologically trivial; see also Appendix~\ref{appendix: WPs in N-DQD model}.

\subsubsection{Antisymmetric SC-coupled double-QD model}\label{sec: ant DQD model}
The normal-coupled double-QD model does not support robust finite-energy WPs and hence does not realize individual-ABS topology. We therefore introduce an SC-coupled double-QD model with a nonlocal interdot pairing $r_{\mathrm{SC}}\neq0$ in addition to normal hopping $r_N\neq0$. The purely SC-coupled limit ($r_N=0$) is also insufficient to produce robust finite-energy WPs; see Appendix~\ref{appendix: WPs in SC-DQD model}. Thus both couplings are required in our working model. Starting from the geometry of Ref.~\cite{Teshler2023}, we allow the central leads $1$ and $2$ to couple to both dots, which generates the phase-dependent nonlocal pairing $r_{\mathrm{SC}}$.

For concreteness, we consider below the antisymmetric geometry with opposite dot energies; see Fig.~\ref{fig: effective models}(c). Dot 1 couples strongly to terminals $0$ and $1$ with amplitude $w_S$ and weakly to terminal $2$ with $w_W<w_S$; dot 2 couples strongly to terminals $2$ and $3$ and weakly to terminal $1$. The two resulting ABSs are therefore centered on the $(0,1)$ and $(2,3)$ terminal pairs but hybridize both through the shared SC terminals and through direct normal hopping. In Eqs.~\eqref{eq: general DQD model} and \eqref{eq: blocks of DQD model}, the parameters are $\varepsilon_1 = -\varepsilon_2 = -\varepsilon$, $\gamma_1 = \Gamma_S(1+e^{i \varphi_1}) + \Gamma_W e^{i \varphi_2} $, $\gamma_2 = \Gamma_S(e^{i\varphi_2}+e^{i \varphi_3}) + \Gamma_W e^{i \varphi_1}$ and $r_{\mathrm{SC}} = \Gamma_{NL}(e^{i\varphi_1}+e^{i\varphi_2})$, with $\Gamma_S=\pi N_0 w_S^2$, $\Gamma_W=\pi N_0 w_W^2$, and $\Gamma_{NL}=\sqrt{\Gamma_S \Gamma_W}$.

Figures~\ref{fig: Chern numbers and WPs}(c) and \ref{fig: Chern numbers and WPs}(f) show the individual-ABS Chern numbers and representative finite-energy WPs in the ABS spectrum, respectively. In Fig.~\ref{fig: Chern numbers and WPs}(f) the two occupied topological ABSs are shown in red and blue, while their particle-hole partners are shown in black. For these parameters, finite-energy WPs generate nontrivial individual-ABS topology while the absence of zero-energy WPs keeps the ground-state Chern number trivial. As we will discuss in Sec.~\ref{sec: Coexistence of ZEWPs and FEWPs}, other parameter choices yield ground-state topology and regimes in which ground-state and individual-ABS topology coexist.

\section{Transconductance: linear response theory}\label{sec: linear response}
In this section, we first recall the zero-temperature quantized transconductance of Ref.~\cite{Riwar2016} and then derive its finite-temperature generalization from the Kubo formula. The resulting expression also separates two ways in which thermal occupations modify the topological response.

In our consideration, we assume quasistatic thermal equilibrium: the ABS relaxation rate is much larger than the Josephson drive frequencies, so the instantaneous occupations are given by the Fermi function. Additionally, we retain only the ABS contribution to the transconductance and neglect continuum states. In the atomic limit used here, the continuum is pushed to infinite energy and its thermal response is therefore negligible.  

\subsection{Zero-temperature transconductance}
Within adiabatic perturbation theory~\cite{Riwar2016}, a slowly varying SC phase results in
\begin{equation}\label{eq: adiabatic}
    I_{\alpha}
    =
    \frac{2e}{\hbar}
    \sum_{E_k <0}
    \frac{\partial E_k}{\partial \varphi_\alpha}
    +
    \frac{4e^2}{\hbar}
    B^{\alpha \beta}V_\beta ,
\end{equation}
where $I_\alpha$ is the SC current in lead $\alpha$ and $V_\beta$ is the voltage applied to lead $\beta$. At $T=0$, the sum runs over the occupied, negative-energy BdG states.

For two incommensurate voltages, the corresponding phases densely cover the synthetic BZ. The first term in Eq.~\eqref{eq: adiabatic} averages to zero, while the Berry-curvature term yields the quantized dc transconductance
\begin{equation}\label{eq: transconductance quantization}
    \overline{I}_\alpha
    =
    G^{\alpha \beta}V_\beta,
    \quad
    G^{\alpha \beta}
    =
    \frac{4 e^2}{h} C^{\alpha \beta},
\end{equation}
where $\overline{I}_\alpha$ denotes the BZ-averaged current. Thus the transconductance between distinct terminals is quantized in units of $4e^2/h$ and is determined by the ground-state Chern number $C^{\alpha\beta}$ \cite{Riwar2016}.

\subsection{Finite-temperature transconductance}
To obtain finite-temperature transconductance, we use linear-response theory. We first define the ABS susceptibility as~\cite{Kurilovich2021,Virtanen2024}
\begin{equation}\label{eq: current}
    I^\alpha(\omega)
    =
    \sum_\beta
    \chi^{ABS}_{\alpha \beta}(\omega)
    \frac{V_\beta(\omega)}{-i \omega}.
\end{equation}
This susceptibility is the ABS-mediated current response in lead $\alpha$ to an ac voltage $V_\beta(\omega)$ applied to lead $\beta$.

In the Lehmann representation, the retarded Kubo response is
\begin{equation}\label{eq: susceptibility}
  \chi^{ABS}_{\alpha \beta}(\omega)
  =
  \sum_{k\neq k'}
  I^\alpha_{kk'}I^\beta_{k'k}
  \frac{
      n_{F}(E_k) - n_F(E_{k'})
  }{
      \hbar \omega +E_k -E_{k'} +i0
  },
\end{equation}
see Appendix~\ref{appendix: ABS susceptibility}. Here $I^\alpha_{kk'} = (2e/
\hbar)
\langle k|\partial_\alpha \hat{H}^{\mathrm{eff}}_{D
}|k' \rangle$ is the current-operator matrix element between BdG eigenstates $|k\rangle$ and $|k'\rangle$, and $n_F(E_k)=1/(e^{E_k/T}+1)$ is the Fermi function (we set the Boltzmann constant $k_B=1$). The sum includes both positive- and negative-energy states, and $+i0$ implies the retarded prescription.

In the adiabatic limit, $\omega\rightarrow0$, we expand the ABS susceptibility to first order in frequency:
\begin{equation}\label{eq: susceptibility expansion}
    \chi^{ABS}_{\alpha \beta}(\omega)
    =
    \chi^{(0)}_{\alpha \beta}
    -
    i \omega \chi_{\alpha \beta}^{(1)}
    +
    O(\omega^2).
\end{equation}
Here $\chi^{(0)}_{\alpha\beta}$ is the static susceptibility and $\chi^{(1)}_{\alpha\beta}$ is the leading dynamical term. Together with the diamagnetic and population contributions, the static response determines the inverse Josephson inductance $L^{-1}_{\alpha\beta}$. Here we focus on the topological term linear in $\omega$:
\begin{equation}\label{eq: chi 1 finite temperature}
    \chi^{(1)}_{\alpha \beta}
    =
    \frac{4e^2}{\hbar}
    B^{\alpha \beta}_T,
\end{equation}
which for $T \neq 0$ defines the finite-temperature Berry curvature
\begin{equation}\label{eq: finite temperature BC}
    B^{ij}_T={\rm Im}\sum_{k \neq k'} \frac{ [n_F(E_k)-n_{F}(E_{k'})] \langle k|\partial_i \hat{H}^{\mathrm{eff}}_{D}|k'\rangle\langle k'|\partial_j\hat{H}^{\mathrm{eff}}_{D}|k\rangle}{(E_k-E_{k'})^2}.
\end{equation}
Averaging over the BZ gives the finite-temperature transconductance
\begin{equation}\label{eq: finite temperature transconductance}
    G_{T}^{\alpha \beta}
    =
    \frac{4e^2}{h}C^{\alpha \beta}_T,
    \quad
    C_T^{\alpha \beta}
    =
    \frac{1}{2\pi}
    \int_0^{2\pi}d\varphi_\alpha
    \int_{0}^{2\pi} d\varphi_{\beta}
    B^{\alpha \beta}_T.
\end{equation}
For compactness, we refer to the BZ integral $C_T^{\alpha\beta}$ as the finite-temperature Chern number. Equation~\eqref{eq: finite temperature transconductance} is then formally identical to the $T=0$ result under $C^{\alpha\beta}\mapsto C_T^{\alpha\beta}$. At $T>0$, however, $C_T^{\alpha\beta}$ is generally noninteger and is not a topological invariant.

At $T=0$, the finite-temperature Berry curvature and finite-temperature Chern number reduce to the Berry curvature and Chern number in Eq.~\eqref{eq: BC and CN}; see Appendix~\ref{appendix: transconductance quantization}. At $T>0$, Fermi factors reweight the ABS contributions, so the finite-temperature and zero-temperature responses no longer coincide. Exact transconductance quantization is therefore lost, and a finite transconductance can also appear where the zero-temperature ground-state Chern number vanishes.

\subsection{Mechanisms of temperature-induced changes in finite-temperature Berry curvature}\label{sec: mechanisms}

Equation~\eqref{eq: finite temperature BC} reveals two distinct thermal contributions to the finite-temperature Berry curvature and, after BZ averaging, to the finite-temperature transconductance. The first (occupied-unoccupied) contribution comes from matrix elements  between the states on opposite sides of the Fermi level at $T=0$. Thermal occupation reduces their Fermi-factor difference, suppressing this contribution. The effect is especially pronounced near zero-energy WPs, where the excitation energy vanishes.

The second (same-sector) contribution  comes from matrix elements between two states in the same $T=0$ occupation sector. Such terms vanish at zero temperature because the two Fermi factors are equal, but become finite when the levels are thermally populated unequally. This mechanism requires at least two positive-energy ABSs (and their particle-hole partners) and is enhanced near finite-energy WPs.    
In a multiband ($M>1$) spectrum both mechanisms generally coexist. To isolate them analytically, we use the single-QD model for the occupied--unoccupied contribution and the four-level Hamiltonian in Eq.~\eqref{eq:FEWP} for the same-sector contribution.

\subsubsection{Single-QD model: occupied--unoccupied contribution}\label{sec: SQD model first mechanism}
For a general single-QD, the $2\times2$ BdG Hamiltonian is determined by a single vector $\mathbf d$ [see Eq.~\eqref{eq: SQD model}]. Equation~\eqref{eq: finite temperature BC} then reduces to
\begin{equation}\label{eq: FTBC in SQD model}
B^{12}_{T} = B^{12} \tanh{\frac{|\mathbf{d}|}{2T}},
\end{equation}
where $|\mathbf d|$ is the phase-dependent ABS energy.

Equation~\eqref{eq: FTBC in SQD model} shows directly that temperature suppresses the Berry-curvature response. The negative-energy state loses occupation while its positive-energy particle-hole partner acquires occupation and contributes with the opposite Berry curvature. Their combined finite-temperature Berry curvature is therefore reduced relative to the $T=0$ Berry curvature.

The suppression is strongly nonuniform in phase space. Regions with $|\mathbf d|\gg T$ are essentially unchanged, whereas low-energy regions with $|\mathbf d|\lesssim T$ are strongly reweighted. In particular, near a zero-energy WP, $|\mathbf d_{\mathrm{ZE}}|\rightarrow0$, the finite-temperature Berry curvature becomes
\begin{equation}
    B^{12}_{T}= - \frac{\delta\varphi_{3}}{4 T} \frac{\det \hat{M}_{ZE}}{|\mathbf{d}_{ZE}|^2}.
\end{equation}
Relative to the zero-temperature Berry curvature, this result is suppressed by $|\mathbf d_{ZE}|/(2T)\rightarrow0$. The finite-temperature Berry curvature nevertheless remains singular at the zero-energy WP. A more useful measure is the finite-temperature Berry-curvature flux through a constant-energy sphere centered on the zero-energy WP:
\begin{equation}
    Q^{|\mathbf{d}_{ZE}|}_T = - Q_{ZE}\tanh \frac{|\mathbf{d}_{ZE}|}{2T} , 
\end{equation}
This flux tends to zero as $|\mathbf d_{ZE}|/T\rightarrow0$. Thus any finite temperature smooths the zero-energy WP contribution after phase averaging: the finite-temperature Chern number becomes continuous and no longer exhibits a quantized jump at the zero-energy WP. 

The single-QD therefore gives a simple phase-resolved picture of thermal suppression. Because the suppression varies across phase space, however, the BZ-integrated finite-temperature Chern number can either decrease or increase depending on how regions of positive and negative Berry curvature are reweighted; see Sec.~\ref{sec: violation of quantization} below. We also note that this monotonic suppression of the finite-temperature Berry curvature is specific to the single-QD model. In multiband systems, different terms in Eq.~\eqref{eq: finite temperature BC} can have opposite signs, so thermal reweighting can either suppress or enhance the finite-temperature Berry curvature.

\subsubsection{Double-QD model: thermally induced same-sector contribution}\label{sec: DQD model second mechanism}
To isolate the same-sector contribution, we use the four-level Hamiltonian from Eq.~\eqref{eq:FEWP}. Beyond the immediate vicinity of a finite-energy WP, it can be viewed as a generic two-ABS model (plus particle-hole partners) parameterized by $E_{\mathrm{FE}}$ and $\mathbf d_{\mathrm{FE}}$. It describes either an individual-ABS topological case, when $|\mathbf d_{\mathrm{FE}}|=0$ somewhere in phase space, or a topologically trivial avoided-crossing case, when $|\mathbf d_{\mathrm{FE}}|\neq0$ everywhere. We assume $E_{\mathrm{FE}}-|\mathbf d_{\mathrm{FE}}|>0$, excluding zero-energy WPs.
 
Because Eq.~\eqref{eq:FEWP} is block diagonal in the two $T=0$ occupation sectors, only same-sector matrix elements contribute, and Eq.~\eqref{eq: finite temperature BC} reduces to
\begin{equation}\label{eq: FTBC FEWP}
B^{12}_T = B^{12}_{\mathrm{FE}} \left(\tanh{\frac{E_{\mathrm{FE}}+|\mathbf{d}_{\mathrm{FE}}|}{2T}} - \tanh{\frac{E_{\mathrm{FE}}-|\mathbf{d}_{\mathrm{FE}}|}{2T}} \right),
\end{equation}
where $B^{12}_{\mathrm{FE}}=-\mathbf d_{\mathrm{FE}}\cdot[\partial_{\varphi_1}\mathbf d_{\mathrm{FE}}\times\partial_{\varphi_2}\mathbf d_{\mathrm{FE}}]/(2|\mathbf d_{\mathrm{FE}}|^3)$ is the Berry curvature of the lowest negative-energy ABS. Unlike Eq.~\eqref{eq: FTBC in SQD model}, this contribution vanishes at $T=0$ and is generated by unequal thermal populations at $T>0$. At fixed phase, it is nonmonotonic in $T$, with asymptotic limits
\begin{equation}
    B^{12}_T = B^{12}_{\mathrm{FE}} \times\begin{cases} 2 e^{(-E_{\mathrm{FE}} + |\mathbf{d}_{\mathrm{FE}}|)/T}, \quad &E_{\mathrm{FE}}, \mathbf{|d_{\mathrm{FE}}|} \gg T, \\
    4( |\mathbf{d}_{\mathrm{FE}}|/T)e^{-E_{\mathrm{FE}}/T}, \quad &E_{\mathrm{FE}} \gg T \gg \mathbf{|d_{\mathrm{FE}}|},\\
    \mathbf{|d_{\mathrm{FE}}|}/T, \quad &E_{\mathrm{FE}}, \mathbf{|d_{\mathrm{FE}}|} \ll T. \\
    \end{cases}
\end{equation}
For the first asymptotic expression, we additionally require the physically relevant condition $E_{\mathrm{FE}}-|\mathbf d_{\mathrm{FE}}|\gg T$. The finite-temperature Berry curvature vanishes both in the low-temperature limit, $E_{FE},|\mathbf d_{\mathrm{FE}}|\gg T$, and in the high-temperature limit, $E_{\mathrm{FE}},|\mathbf d_{\mathrm{FE}}|\ll T$, where all four occupations approach one half.

Equation~\eqref{eq: FTBC FEWP} applies to both topological crossings and trivial avoided crossings. At the same time, near a finite-energy WP in the individual-ABS topological regime, $|\mathbf{d}_{\mathrm{FE}}| \rightarrow0$, substituting the linearized parameters from Eq.~\eqref{eq:FEWP_block} gives 
\begin{equation}\label{eq: FTBC of FEWP}
     B_T^{12}=-\frac{\delta\varphi_3\det\hat M_{\mathrm{FE}}}{2T|\mathbf d_{\mathrm{FE}}|^2\cosh^2(E_{\mathrm{FE}}/2T)},
\end{equation}
The expression scales as $(1/T)\exp(-E_{\mathrm{FE}}/T)$ for $T/E_{\mathrm{FE}}\ll1$ and as $1/T$ for $T/E_{\mathrm{FE}}\gg1$. As for a zero-energy WP, the finite-temperature Berry curvature is singular at the crossing, although the thermally active region shrinks at low temperature. The corresponding finite-temperature Berry-curvature flux through an energy sphere centered on the finite-energy WP is finite:

\begin{equation}
    Q^{\mathbf{|d_{\mathrm{FE}}|}}_{T} = -Q_{\mathrm{FE}} \left(\tanh{\frac{E_{\mathrm{FE}}+|\mathbf{d}_{\mathrm{FE}}|}{2T}} - \tanh{\frac{E_{\mathrm{FE}}-|\mathbf{d}_{\mathrm{FE}}|}{2T}} \right),
\end{equation}
where, for simplicity, we consider an untilted finite-energy WP. The finite-temperature Berry-curvature flux vanishes in both the low- and high-temperature limits. Consequently, the finite-temperature Chern number has no quantized jump associated with a finite-energy WP at any temperature.

These results show that same-sector thermal reweighting can generate a finite-temperature Berry curvature that is strictly absent at $T=0$. Because this response is highly nonuniform in phase space, however, the corresponding finite-temperature Chern number must be evaluated numerically.

\section{Violation of the transconductance quantization}\label{sec: violation of quantization}

In this section, we investigate the thermal breakdown of transconductance quantization in four-terminal MTJJs. In terms of the finite-temperature Chern number, the breakdown manifests itself as
\begin{equation}
    C^{12}_T \neq C^{12}, \quad T\neq 0.
\end{equation}
For the QD models considered here, the breakdown is dominated by thermal suppression of the occupied--unoccupied contribution.

\begin{figure*}[t]
\includegraphics[width=0.33\columnwidth]{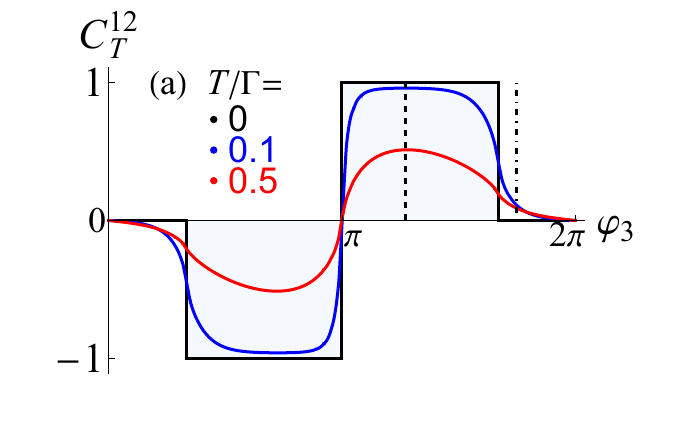}\centering \hfill
\includegraphics[width=0.33\columnwidth]{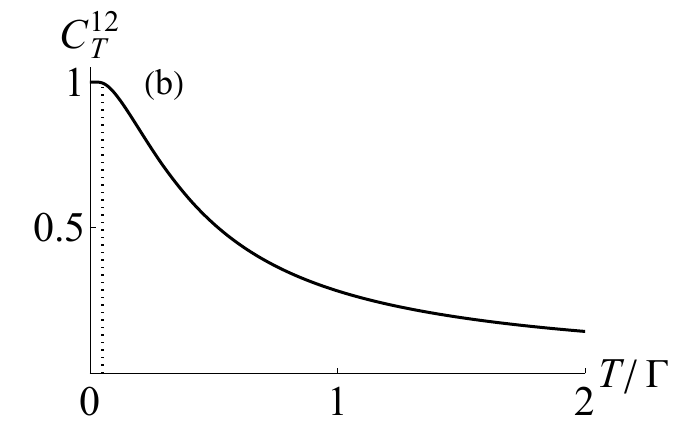}\centering \hfill
\includegraphics[width=0.33\columnwidth]{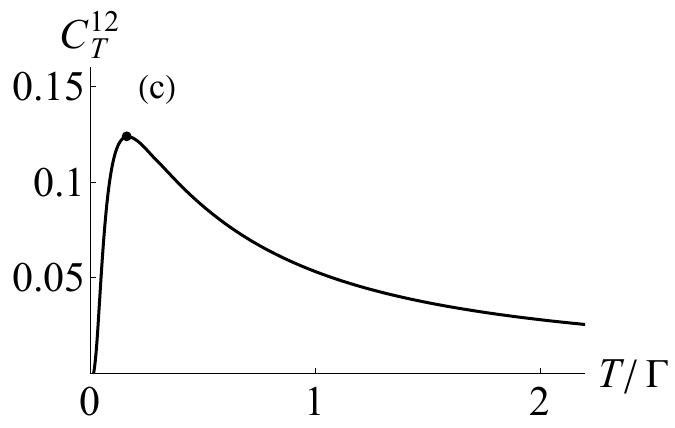}\centering
\\
\includegraphics[width=0.33\columnwidth]{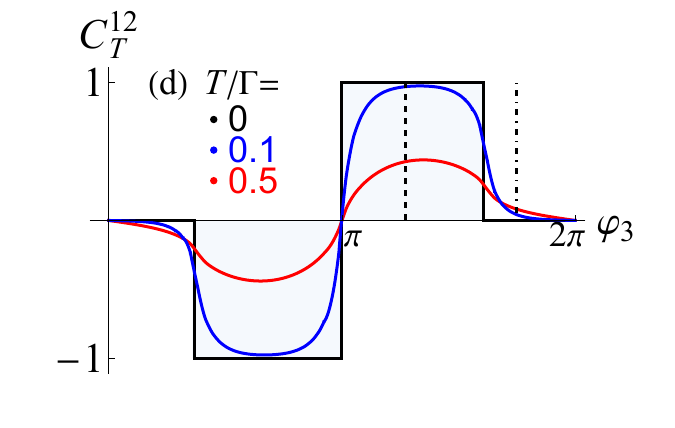}\centering \hfill
\includegraphics[width=0.33\columnwidth]{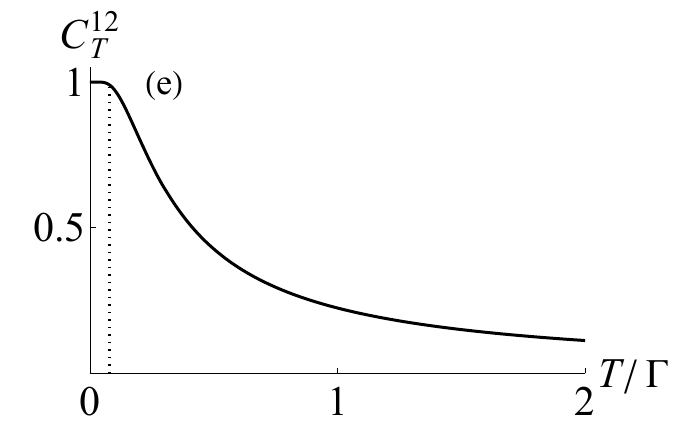}\centering \hfill
\includegraphics[width=0.33\columnwidth]{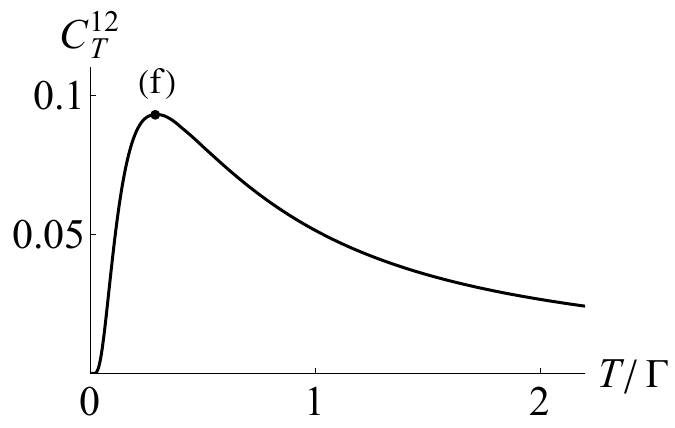}\centering
\caption{Finite-temperature Chern number $C_T^{12}$ for the symmetric single-QD (top row) and symmetric normal-coupled double-QD (bottom row). Panels (a) and (d) show $C_T^{12}(\varphi_3)$ at the indicated temperatures; light-blue shading marks the ground-state topological regions. Parameters are $\varepsilon/\Gamma=-0.2$, $\tau=0.1$ (top) and $\varepsilon/\Gamma=1$, $r_N/\Gamma=1.5$ (bottom). At $T=0$, $C_T^{12}=C^{12}$; finite temperature smooths the zero-energy-WP steps and suppresses the plateau. Dashed lines at $\varphi_3=4$ in (a,d) correspond to the topological cuts in (b,e), while dash-dotted lines at $\varphi_3=5.5$ correspond to the trivial cuts in (c,f). Dotted lines in (a,d) delimit the nearly quantized regions. The topological cuts decrease monotonically with $T$; the trivial cuts are nonmonotonic, with maxima marked by black dots.}
\label{fig: finite temperature Chern number}
\end{figure*}

We evaluate Eqs.~\eqref{eq: finite temperature BC} and \eqref{eq: finite temperature transconductance} numerically for the symmetric single-QD and normal-coupled double-QD models. Figures~\ref{fig: finite temperature Chern number}(a) and \ref{fig: finite temperature Chern number}(d) show $C_T^{12}(\varphi_3)$ in these cases, respectively. Both models behave similarly: at low $T$, deviations from quantization are small except near zero-energy WPs, where the excitation gap closes and even weak thermal occupation smooths the response. As $T$ increases, the deviation extends throughout the whole ground-state topological region.

Figures~\ref{fig: finite temperature Chern number}(b),(c) and (e),(f) show $C_T^{12}(T)$ at fixed $\varphi_3$ for the symmetric single-QD and normal-coupled double-QD models, respectively. In a ground-state topological region [Figs.~\ref{fig: finite temperature Chern number}(b) and (e)  which correspond to the dashed vertical lines in Figs.~\ref{fig: finite temperature Chern number}(a) and (d), respectively], $C_T^{12}$ decreases monotonically because thermal occupation suppresses the dominant Berry curvature contribution; Appendix~\ref{appendix: Visualization of BC for SQD and N-DQD models} shows the corresponding finite-temperature Berry-curvature maps. In a ground-state trivial region [Figs.~\ref{fig: finite temperature Chern number}(c) and (f)  which correspond to the dashed-dotted vertical lines in Figs.~\ref{fig: finite temperature Chern number}(a) and (d), respectively], positive- and negative-Berry-curvature contributions cancel at $T=0$ but are thermally reweighted at different rates. The resulting finite-temperature transconductance is finite and nonmonotonic and vanishes again at high temperature as the ABS occupations become equal.

\section{Finite-temperature detection of individual-ABS topology}\label{sec: Detection of FEWPs}

We next ask whether finite-temperature transconductance can reveal individual-ABS topology that is invisible in the ground-state Chern number. We use the antisymmetric SC-coupled double-QD model in a regime with finite-energy WPs but no zero-energy WPs, so $C^{12}=0$ although individual ABSs carry nonzero Chern numbers; the response is therefore governed by the thermally induced same-sector contribution. We then compare this regime with a topologically trivial state close to the topological transition to assess how specifically the finite-temperature transconductance identifies individual-ABS topology.

\subsection{Finite-temperature Chern number of the individual-ABS topological state}

Using Eqs.~\eqref{eq: finite temperature BC} and \eqref{eq: finite temperature transconductance}, we compute $C_T^{12}(\varphi_3)$ for the antisymmetric SC-coupled double-QD model. Figure~\ref{fig: FTCN in the SC-DQD model}(a) shows finite-temperature peaks precisely in the interval where the individual ABSs carry nonzero Chern numbers. The underlying finite-temperature Berry curvature is thermally generated and its phase-space maps are shown in Appendix~\ref{appendix: Visualization of BC in antisymmetric SC-DQD model}.

At fixed $\varphi_3$ inside this interval, Fig.~\ref{fig: FTCN in the SC-DQD model}(b) shows a nonmonotonic $C_T^{12}(T)$ with a maximum at finite temperature. Unlike the trivial-region maxima in Figs.~\ref{fig: finite temperature Chern number}(c) and (f), this maximum is not caused by unequal suppression of pre-existing regions of positive and negative Berry curvature. Instead, same-sector thermal reweighting generates finite-temperature Berry curvature with nonuniform phase-space weight; see  Appendix~\ref{appendix: Visualization of BC in antisymmetric SC-DQD model}. Thus individual-ABS topology produces two characteristic signatures in this model: peaks in $C_T^{12}(\varphi_3)$ and a nonmonotonic $C_T^{12}(T)$.

\subsection{Individual-ABS topology vs. trivial topology}\label{sec: FEWPs vs trivial topology}

Thermally induced finite-temperature transconductance features do not by themselves imply nontrivial topology. To test their specificity, we compare them with a topologically trivial system that nevertheless has a nonzero phase-resolved Berry curvature distribution.
 
\begin{figure*}[t]
\centering
\includegraphics[width=0.4\columnwidth]{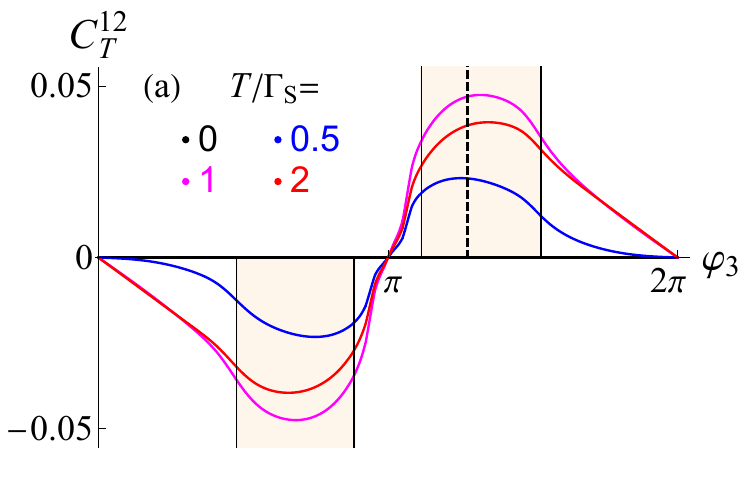} \hspace{1 cm}
\includegraphics[width=0.4\columnwidth]{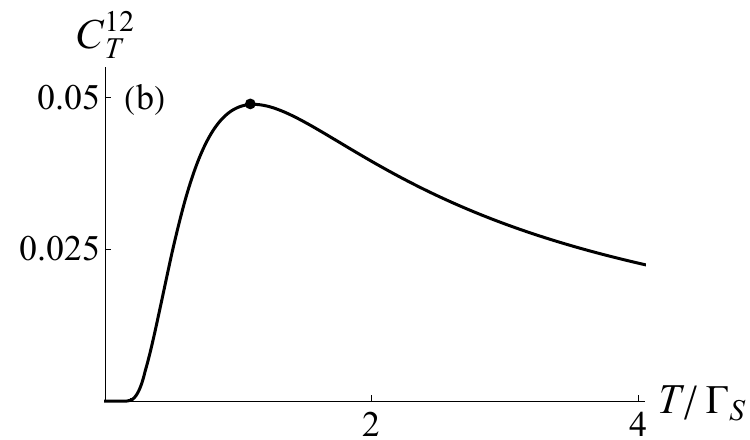} \\
\includegraphics[width=0.4\columnwidth]{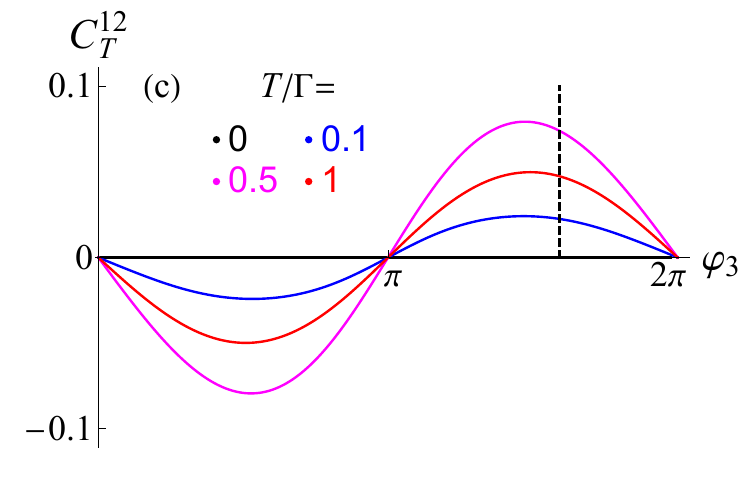} \hspace{1 cm}
\includegraphics[width=0.4 \columnwidth]{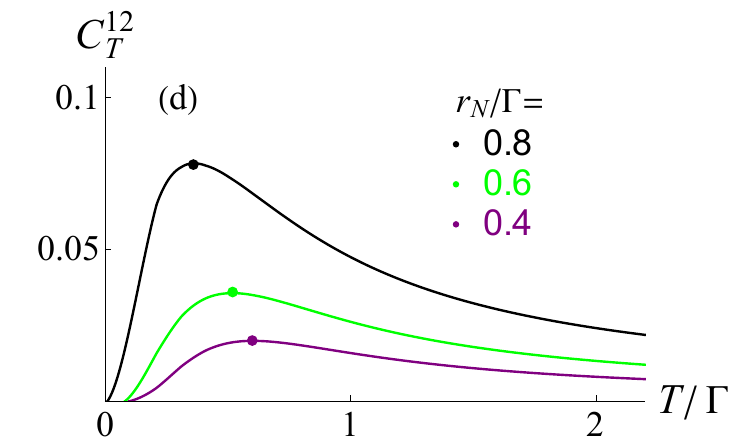} \\
\caption{Finite-temperature Chern number $C_T^{12}$ in an individual-ABS topological regime (top row) and a trivial regime (bottom row). Panels (a,b) correspond to the antisymmetric SC-coupled double-QD with $\varepsilon/\Gamma_S=1.2$, $\Gamma_W/\Gamma_S=0.3$, and $r_N/\Gamma_S=0.1$. In (a), the light-red interval marks nonzero individual-ABS Chern numbers; vertical lines indicate its boundaries and the dashed line at $\varphi_3=4$ defines the cut used in (b). Thermally induced peaks appear in $C_T^{12}(\varphi_3)$ in (a), with the corresponding nonmonotonic $C_T^{12}(T)$ shown in (b). Panels (c,d) correspond to the symmetric normal-coupled double-QD in the trivial regime with $\varepsilon/\Gamma=1$ and $r_N/\Gamma=0.8$ in (c). The dashed line at $\varphi_3=5$ defines the cut used in (d), where $C_T^{12}(T)$ is shown for several $r_N$. Similar thermal peaks appear in (c), with a nonmonotonic $C_T^{12}(T)$ dependence in (d); the peak magnitude decreases as the system is tuned away from the transition at $r_N/\Gamma=1$.}
\label{fig: FTCN in the SC-DQD model}
\end{figure*}

For comparison, we choose the symmetric normal-coupled double-QD model on the trivial side of its topological transition, $r_N/\Gamma<1$~\cite{Teshler2023}. Figures~\ref{fig: FTCN in the SC-DQD model}(c) and (d) again show phase-dependent peaks of $C^{12}_T(\varphi_3)$ and a nonmonotonic $C_T^{12}(T)$, although all individual-ABS Chern numbers vanish. Appendix~\ref{appendix: Visualization of FTBC in trivial model} shows the corresponding finite-temperature Berry-curvature maps. These peaks are generally weaker and decrease rapidly as $r_N$ is tuned farther into the trivial side of the transition, becoming negligible sufficiently far from the phase boundary.

Pronounced finite-temperature transconductance peaks are therefore not a unique diagnostic of nonzero individual-ABS Chern numbers. Phase-resolved measurements of the finite-temperature Berry curvature can, however, distinguish the two cases in these simple QD models despite their similar averaged finite-temperature transconductance; see Appendices \ref{appendix: Visualization of BC in antisymmetric SC-DQD model} and \ref{appendix: Visualization of FTBC in trivial model}.

\section{Coexistence of ground-state and individual-ABS topology}\label{sec: Coexistence of ZEWPs and FEWPs}

Section~\ref{sec: violation of quantization} showed the thermal suppression of finite-temperature transconductance associated with ground-state topology (zero-energy WPs), whereas Sec.~\ref{sec: Detection of FEWPs} showed thermally generated signatures of individual-ABS topology (finite-energy WPs). We now consider the antisymmetric SC-coupled double-QD in regimes where both types of WPs coexist.

\begin{figure*}[t]
\includegraphics[width=0.33\columnwidth]{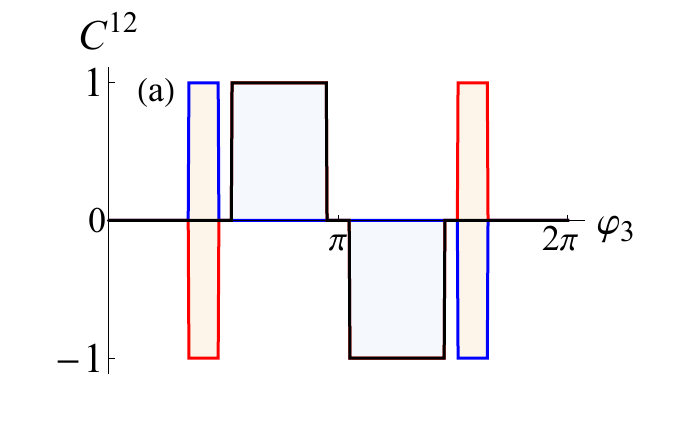}\centering \hfill
\includegraphics[width=0.33\columnwidth]{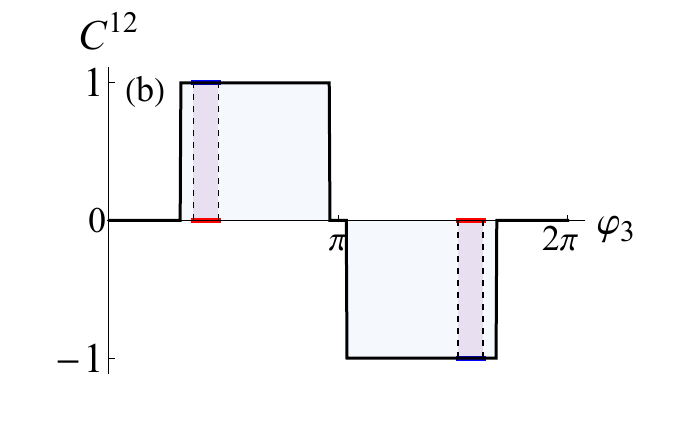}\centering \hfill
\includegraphics[width=0.33\columnwidth]{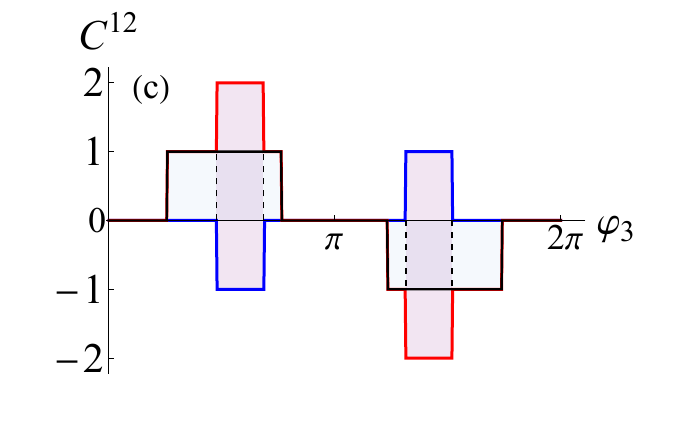}\centering\\
\includegraphics[width=0.33\columnwidth]{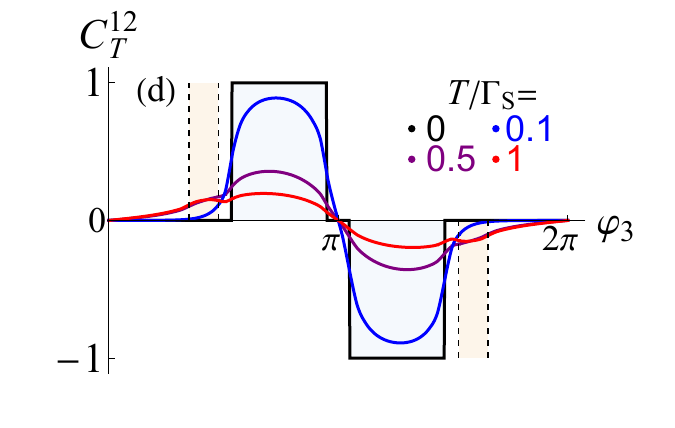}\centering 
\hfill
\includegraphics[width=0.33\columnwidth]{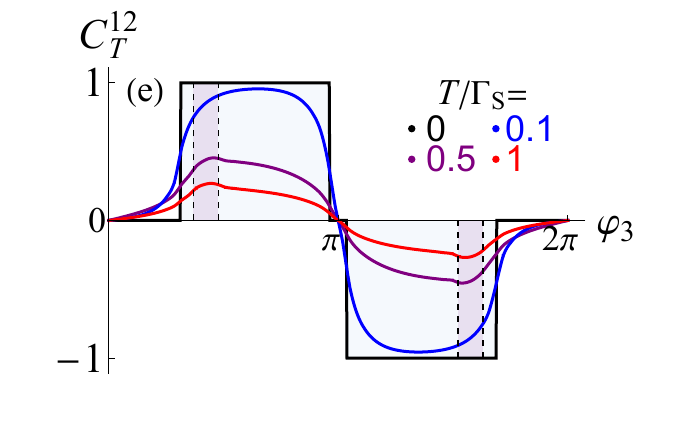}\centering 
\hfill
\includegraphics[width=0.33\columnwidth]{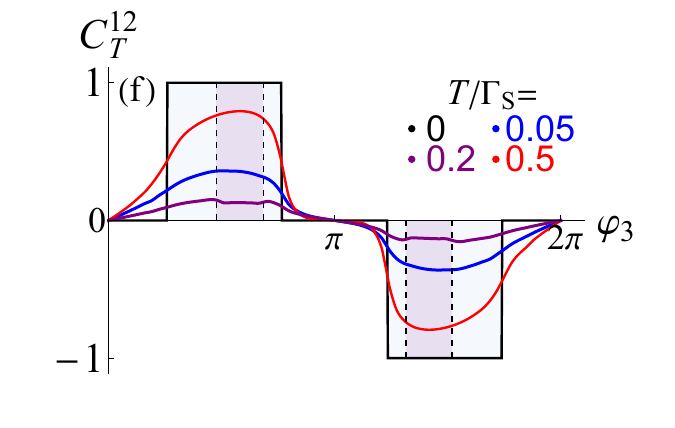}\centering
\caption{Coexisting ground-state and individual-ABS topology in the antisymmetric SC-coupled double-QD model with $\Gamma_W/\Gamma_S=0.3$. Ground-state, individual-ABS, and overlapping topological regions are shaded light blue, light red, and light purple, respectively; dashed lines mark the boundaries of the individual-ABS topological regions. Panels (a)--(c) show $C^{12}$ (black), $C_1^{12}$ (blue), and $C_2^{12}$ (red), while panels (d)--(f) show $C_T^{12}(\varphi_3)$. Panels (a,d) use $\varepsilon/\Gamma_S=0.1$ and  $r_N/\Gamma_S=1.2$; panels (b,e), $\varepsilon/\Gamma_S=0.1$ and  $r_N/\Gamma_S=0.6$; and panels (c,f), $\varepsilon/\Gamma_S=0.2$ and $r_N/\Gamma_S=0.1$. In (a,d), the two topological intervals are separated; in (b,e) and (c,f), they overlap. Thermal individual-ABS contributions appear separately in (d), as a peak on the suppressed ground-state response in (e), and as a dip in (f).}
\label{fig: Coexistence of WPs and FEWPs}
\end{figure*}

Figures~\ref{fig: Coexistence of WPs and FEWPs}(a)--(c) show the ground-state and individual-ABS Chern numbers for three parameter sets. The ground-state and individual-ABS topological intervals are either separated [Fig.~\ref{fig: Coexistence of WPs and FEWPs}(a)] or overlap [Figs.~\ref{fig: Coexistence of WPs and FEWPs}(b) and (c)], with different individual-ABS Chern number patterns in the two overlapping cases.

The corresponding finite-temperature Chern numbers are shown in Figs.~\ref{fig: Coexistence of WPs and FEWPs}(d)--\ref{fig: Coexistence of WPs and FEWPs}(f). When the intervals are separated [Fig.~\ref{fig: Coexistence of WPs and FEWPs}(d)], the two thermal signatures are also separated: $|C_T^{12}|$ decreases in the ground-state topological interval, while finite-temperature peaks appear in the individual-ABS topological interval. When the intervals overlap [Figs.~\ref{fig: Coexistence of WPs and FEWPs}(e) and (f)], these trends superpose. Depending on the ordering of the individual-ABS Chern numbers, the finite-energy-WP-related contribution appears as either a peak [Fig.~\ref{fig: Coexistence of WPs and FEWPs}(e)] or a dip [Fig.~\ref{fig: Coexistence of WPs and FEWPs}(f)] on top of the overall thermal suppression of the ground-state response. 

These results show that the finite-temperature response can be understood as a superposition of two physically distinct mechanisms: suppression of the contribution associated with ground-state topology and generation of a contribution associated with individual-ABS \mbox{topology}. Their relative sign and phase-space overlap determine whether the latter appears as a peak or a dip in the total response.

\section{Discussion}\label{sec:discussion}

\subsection{Strength of the temperature-induced effects}
The preceding sections identified both phase-resolved (Sec.~\ref{sec: mechanisms}) and phase-averaged (Secs.~\ref{sec: violation of quantization}--\ref{sec: Coexistence of ZEWPs and FEWPs}) thermal effects. We now discuss their magnitude.

Temperature can strongly modify the  phase-resolved response. In the single-QD model, $|\mathbf d|\ll T$ strongly suppresses the finite-temperature Berry curvature, whereas near a finite-energy WP even a small temperature can generate a large finite-temperature Berry curvature. The phase-resolved response can therefore show either pronounced suppression or enhancement, see Appendix~\ref{sec: Visualization of finite-temperature Berry Curvature}.

The BZ-averaged effect can also be large, as the violation of quantization in Sec.~\ref{sec: violation of quantization} demonstrates. The thermally generated signal associated with individual-ABS topology is smaller in our examples: its typical maximum is $C_T^{12}\sim0.1$. This is not a fundamental bound; it reflects the detailed ABS dispersion of the chosen model. A larger response would require a broad phase-space region in which one relevant ABS is thermally active while the other remains effectively unoccupied. In this case, the finite-temperature Chern number could reach $C^{12}_T \rightarrow1$.

\subsection{Reconstruction of the individual-ABS Chern numbers from the finite-temperature transconductance}

The experimentally relevant inverse problem is whether finite-temperature transconductance measurements can reconstruct ABS topology. At $T=0$, the averaged transconductance directly measures the ground-state Chern number~\cite{Riwar2016}. At low temperature and sufficiently far from a zero-energy WP, this quantization remains a good approximation. At higher temperature, the sharp Chern number steps are smoothed, but their locations can still be inferred from the crossover of $C_T^{12}(\varphi_3)$ and from its characteristic temperature dependence.

Reconstructing individual-ABS Chern numbers is less direct. As shown in Sec.~\ref{sec: FEWPs vs trivial topology}, thermally induced peaks also occur on the trivial side near a topological transition. Their amplitude decreases as the system is tuned away from the transition, so a pronounced finite-temperature transconductance feature signals either proximity to a transition or topological ABS bands but does not uniquely determine the individual-ABS Chern numbers.

The ambiguity increases when several resonant ABSs contribute. In the antisymmetric SC-coupled double-QD model, zero-energy WPs and finite-energy WPs remain distinguishable because their phase intervals and temperature dependences can be tracked separately (Sec.~\ref{sec: Coexistence of ZEWPs and FEWPs}), whereas generic multiband junctions, in principle, may produce similar finite-temperature transconductance curves from different microscopic spectra. Finite-temperature transconductance should therefore be viewed as a sensitive probe: it robustly identifies ground-state topology and highlights phase regions consistent with nontrivial individual-ABS structure, but additional information is required for an unambiguous individual-ABS topological assignment.

\subsection{Alternative measurement protocols}

Because the BZ-averaged finite-temperature transconductance does not uniquely determine individual-ABS Chern numbers, it is most useful in combination with phase-resolved and spectroscopic probes. The finite-temperature Berry-curvature maps in Appendices~\ref{appendix: Visualization of BC in antisymmetric SC-DQD model} and \ref{appendix: Visualization of FTBC in trivial model} distinguish, for the present models, thermally generated same-sector response which is a characteristic thermal feature of the individual-ABS topology from unequal suppression of pre-existing Berry curvature which is a characteristic thermal feature of the trivial or ground-state topology. However, this distinction need not remain unique in a more complicated multiband case. Finite-temperature transconductance can therefore first identify promising control-phase intervals. Tunneling spectroscopy~\cite{Antonelli2025,Jung2025} can then test for finite-energy ABS crossings, while microwave spectroscopy provides a complementary check because the transitions at $\omega=2E_1$, $2E_2$, and $E_1+E_2$ merge at $E_1=E_2$. The protocols of Refs.~\cite{Klees2020,Klees2021} can additionally reconstruct the Berry curvature of a selected ABS from transition matrix elements, providing an independent test of the individual-ABS topology.

\section{Conclusions}\label{sec:conclusions}

We studied the finite-temperature transconductance of four-terminal MTJJs with topological Andreev spectra. Their synthetic topology is controlled by WPs in SC-phase space: zero-energy WPs change the ground-state Chern number, whereas finite-energy WPs redistribute Chern numbers among individual ABSs without changing the ground-state sum. Using effective single-QD and double-QD models, we identified regimes with zero-energy WPs, finite-energy WPs, and their coexistence; the antisymmetric SC-coupled double-QD with both normal hopping and nonlocal SC coupling is the minimal model used here to realize robust finite-energy crossings and nontrivial individual-ABS Chern numbers.

We derived the finite-temperature transconductance using linear-response theory. Thermal occupation has two consequences: it suppresses contributions connecting states on opposite sides of the Fermi level and activates contributions between states in the same $T=0$ occupation sector. The first mechanism manifests itself in the breakdown of the exact transconductance quantization, while the second generates a finite response near finite-energy WPs even when the $T=0$ ground-state Chern number is zero.

For the single-QD and normal-coupled double-QD models, quantization remains accurate when $T$ is small compared with the relevant ABS energies, except near zero-energy WPs where the gap closes. In ground-state topological regions, $|C_T^{12}|$ decreases with temperature; in trivial regions, unequal thermal reweighting can produce a nonmonotonic finite response.

For individual-ABS topology, the antisymmetric SC-coupled double-QD develops finite-temperature peaks in $C_T^{12}$ within the phase interval bounded by finite-energy WPs, and a nonmonotonic temperature dependence. Similar features can arise in a topologically trivial model close to a transition, so the BZ-averaged finite-temperature transconductance is not by itself a unique individual-ABS topology diagnostic. At the same time, phase-resolved finite-temperature response measurements can help distinguish between these two scenarios.

When zero-energy WPs and finite-energy WPs coexist, their thermal signatures superpose as suppression of the ground-state response together with peaks or dips associated with individual-ABS topology. These results establish finite temperature both as a limitation on quantized transconductance and as a practical source of otherwise hidden information about the topology of Andreev spectra.

\section*{Acknowledgements}
We thank Daniel Bobok, Gustavo Diniz, Joren Harms, Michael Hein, Tom\'a\v{s} Novotn\'y,  Aleksandr Svetogorov, Peter Zalom, and Martin \v{Z}onda for useful discussions. 

\paragraph{Funding information}
We acknowledge support by the DAAD PPP (Projekt-ID: 57753337) and the Deutsche Forschungsgemeinschaft (DFG; German Research Foundation) by SFB 1432 (Project No. 425217212).

\begin{appendix}
\numberwithin{equation}{section}

\section{Effective Hamiltonians of QD models}\label{appendix: effective Hamiltonian}

\subsection{Derivation scheme}
\label{appendix: derivation of low energy Hamiltonian}

\begin{figure}[t]
\centering
\includegraphics[width=0.5\columnwidth]{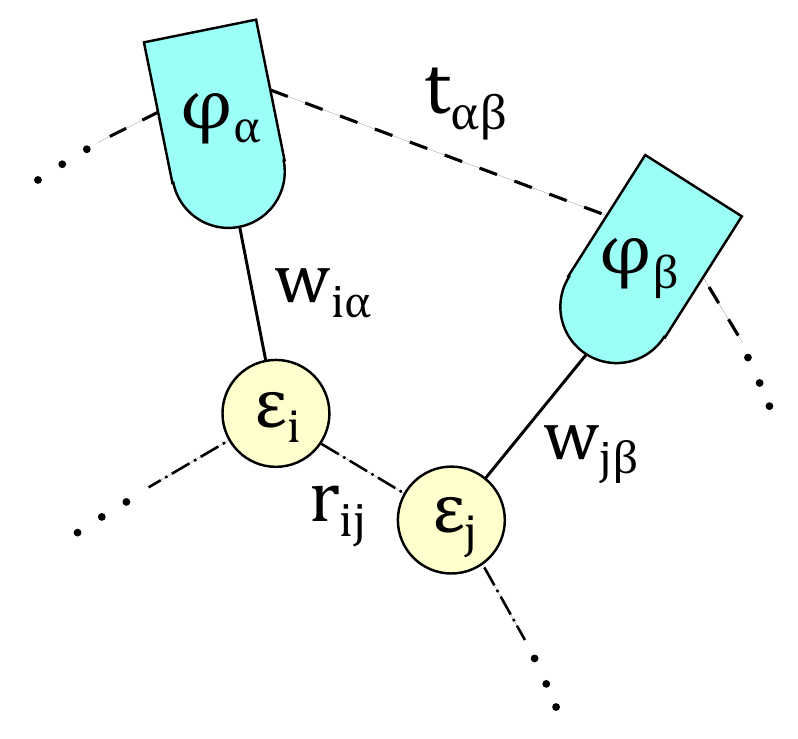}
\caption{Generic effective QD model for an MTJJ. The $N$ SC leads have phases $\varphi_\alpha$ ($\alpha=0,\ldots,N-1$), while the $M$ resonant ABSs are represented by single-level QDs with on-site energies $\varepsilon_i$ ($i=1,\ldots,M$). Dot $i$ couples to lead $\alpha$ with amplitude $w_{i\alpha}$; dots $i$ and $j$ may be coupled by normal hopping $r_{ij}$, and SC leads $\alpha$ and $\beta$ may be directly coupled by $t_{\alpha\beta}$.}
\label{fig: QD model}
\end{figure}

We consider $M$ single-level QDs, indexed by $i,j=1,\ldots,M$, coupled to $N$ SC leads, indexed by $\alpha,\beta=0,\ldots,N-1$; see Fig.~\ref{fig: QD model}. We use the Green's function technique to integrate out SC leads and derive the effective BdG Hamiltonian for the resonant ABSs \cite{Klees2020,Klees2021, Teshler2023}. 

In the Nambu representation, the full BdG Hamiltonian has the block form
\begin{equation}
    \mathcal H_{\mathrm{BdG}} =
    \begin{pmatrix}
        \hat H_D & \hat H_{DS} \\
        \hat H_{DS}^\dagger & \hat H_S
    \end{pmatrix},
    \label{eq:BdG_app}
\end{equation}
where $\hat H_D$ is $2M\times2M$ dot block, $\hat H_S$ is $2N\times2N$ SC lead block, and $\hat H_{DS}$ is $2M\times2N$ dot-leads block. We use $\hat\tau_i$ for Pauli matrices in Nambu space. The blocks are given by
\begin{align}
    \left(\hat H_D\right)_{ij}
    &=\left(\varepsilon_i\delta_{ij}+r_{ij}\right)\hat\tau_3,\label{eq:HD_BdG_app}\\
    \left(\hat H_{S0}\right)_{\alpha\beta}
    &=\left[\xi_{\mathbf k}\hat\tau_3+\Delta_\alpha e^{i\varphi_\alpha\hat\tau_3}\hat\tau_1\right]\delta_{\alpha\beta},\label{eq:HS0_BdG_app}\\
    \left(\hat H_{S-S}\right)_{\alpha\beta}
    &=t_{\alpha\beta}\hat\tau_3,\label{eq:HSS_BdG_app}\\
    \left(\hat H_{DS}\right)_{i\alpha}
    &=w_{i\alpha}\hat\tau_3.\label{eq:HDS_BdG_app}
\end{align}
Here $\xi_{\mathbf k}=\hbar^2k^2/(2m)-E_F$, $\varepsilon_i$ is on-site dot energy, $r_{ij}$ is the normal interdot hopping, $t_{\alpha\beta}$ is direct lead--lead hopping, and $w_{i\alpha}$ is the dot--lead tunneling amplitude; see Fig.~\ref{fig: QD model}. We assume real, spin-preserving tunneling amplitudes.

Integrating out the SC leads gives the dot self-energy
\begin{equation}
    \hat\Sigma_D(E)=\hat H_{DS}\hat g_{SS}(E)\hat H_{DS}^\dagger,
    \label{eq:self_energy_app}
\end{equation}
where, to first order in weak direct lead--lead tunneling,
\begin{equation}
    \hat g_{SS}(E)\simeq\hat g_S(E)+\hat g_S(E)\hat H_{S-S}\hat g_S(E).
    \label{eq:gSS_expansion_app}
\end{equation}
In the atomic limit, $E\ll\Delta_\alpha$, the Green function of an isolated SC lead reduces to
\begin{equation}
    \left(\hat g_S\right)_{\alpha\beta}
    =-\pi N_0 e^{i\varphi_\alpha\hat\tau_3}\hat\tau_1\delta_{\alpha\beta}.
    \label{eq:gS_atomic_app}
\end{equation}
The self-energy is then energy independent, and the effective dot Hamiltonian is
\begin{equation}
    \hat H_D^{\mathrm{eff}}=\hat H_D+\hat\Sigma_D.
    \label{eq:Heff_app}
\end{equation}
It contains the bare dot levels and hoppings together with proximity-induced pairing generated by virtual tunneling through the SC leads.

\subsection{Effective Hamiltonians of the single-QD and double-QD models}\label{appendix: Different QD models}

We now apply Appendix~\ref{appendix: derivation of low energy Hamiltonian} to the single-QD, normal-coupled double-QD, and SC-coupled double-QD models. The resulting effective Hamiltonians are summarized below.

\subsubsection{Single-QD model}
For the single-QD, one dot couples to four SC terminals, and neighboring terminals are directly coupled. Hence $t_{\alpha\beta}\neq0$ only for $\beta=\alpha\pm1$, with terminal indices understood modulo four. Keeping both lead--lead and dot--lead tunneling in the self-energy yields 
\begin{gather}
    \hat H_D^{\mathrm{eff}}=\mathbf d\cdot\hat{\boldsymbol\tau}, \notag \\
    \mathbf{d} =
    \begin{pmatrix}
        \sum^3_{\alpha=0} \Gamma _{\alpha} \cos \varphi_\alpha \\
        - \sum^3_{\alpha=0} \Gamma_\alpha \sin \varphi_\alpha \\
        \varepsilon_1
        - 2\sum^3_{\alpha=0}
        \tau_{\alpha(\alpha+1)} \Gamma_{\alpha(\alpha+1)}
        \cos (\varphi_\alpha - \varphi_{\alpha+1})
    \end{pmatrix}.
    \label{eq: 1D system}
\end{gather}
Here $\Gamma_\alpha= \pi N_0w_{1\alpha}^2$, $\tau_{\alpha,\alpha+1}=\pi N_0t_{\alpha,\alpha+1}$, and $\Gamma_{\alpha(\alpha+1)}=\pi N_0w_{1\alpha}w_{1(\alpha+1)}$. Equation~\eqref{eq: 1D system} reduces to Eq.~\eqref{eq: symmetric SQD model} when $t_{\alpha(\alpha+1)}=t_{\alpha(\alpha-1)}=t$ and $w_{1\alpha}=w$ for all $\alpha=0,\ldots,3$.

\subsubsection{Normal-coupled double-QD model}
The normal-coupled double-QD contains two dots: dot 1 couples to terminals $0$ and $1$, and dot 2 to terminals $2$ and $3$. The dots are directly coupled by $r_{12}=r_N$. As in Ref.~\cite{Teshler2023}, the dot--lead self-energy gives the block Hamiltonian of Eq.~\eqref{eq: general DQD model}. In the general case, $\varepsilon_1\neq\varepsilon_2$, $\gamma_1=|\gamma_1|e^{i\psi_1}=\Gamma_{10}+\Gamma_{11}e^{i\varphi_1}$, and $\gamma_2=|\gamma_2|e^{i\psi_2}=\Gamma_{22}e^{i\varphi_2}+\Gamma_{23}e^{i\varphi_3}$, with $\Gamma_{i\alpha}=\pi N_0w_{i\alpha}^2$. The symmetric normal-coupled double-QD follows for $\varepsilon_1=\varepsilon_2=\varepsilon$ and $\Gamma_{10}=\Gamma_{11}=\Gamma_{22}=\Gamma_{23}=\Gamma$.

\subsubsection{SC-coupled double-QD model}
The SC-coupled double-QD model extends the normal-coupled double-QD by allowing the central terminals $1$ and $2$ to couple to both dots. The shared leads generate a phase-dependent nonlocal SC coupling $r_{\mathrm{SC}}$ in addition to the normal hopping $r_N$. The effective Hamiltonian retains the block structure of Eq.~\eqref{eq: general DQD model}, with $r_{\mathrm{SC}}=|r_{\mathrm{SC}}|e^{i\psi_{\mathrm{SC}}}=\Gamma_{NL}^{(1)}e^{i\varphi_1}+\Gamma_{NL}^{(2)}e^{i\varphi_2}$, where $\Gamma_{NL}^{(1)}=\pi N_0w_{11}w_{21}$ and $\Gamma_{NL}^{(2)}=\pi N_0w_{12}w_{22}$. The local pairings are $\gamma_1=\Gamma_{10}+\Gamma_{11}e^{i\varphi_1}+\Gamma_{12}e^{i\varphi_2}$ and $\gamma_2=\Gamma_{21}e^{i\varphi_1}+\Gamma_{22}e^{i\varphi_2}+\Gamma_{23}e^{i\varphi_3}$ in Eq.~\eqref{eq: blocks of DQD model}. The antisymmetric model of the main text follows for $\varepsilon_1=-\varepsilon_2=-\varepsilon$, $w_{10}=w_{11}=w_{22}=w_{23}=w_S$, and $w_{12}=w_{21}=w_W$, with $\Gamma_S=\pi N_0w_S^2$ and $\Gamma_W=\pi N_0w_W^2$.

\section{WPs in different QD models}\label{appendix: conditions for WPs}
This appendix summarizes the conditions for zero- and finite-energy WPs in the single-QD and double-QD models.

\subsection{Zero-energy WPs in the single-QD model}\label{appendix: ZEWPs in SQD model}
With only one positive-energy ABS, the single-QD can host zero-energy WPs but not finite-energy WPs. Zero-energy WPs require $\mathbf d=0$, i.e., three simultaneous trigonometric conditions $d_x=d_y=d_z=0$. For a generic single-QD these conditions are solved numerically; in the symmetric model, symmetry allows an analytic criterion~\cite{Klees2020}. The result is $-8<\varepsilon/(\tau \Gamma) < 0$. 

\subsection{WPs in double-QD models}\label{appendix: WPs in DQD model}
Before specializing the double-QD model, we summarize the degeneracy conditions for a generic $4\times4$ double-QD BdG Hamiltonian.

A double-QD has a $4\times4$ BdG Hamiltonian with two particle-hole pairs, $\pm E_1^{ABS}$ and $\pm E_2^{ABS}$, so its characteristic polynomial can be written as
\begin{equation}\label{eq: det}
    \mathrm{det}(\hat{H}^{\mathrm{eff}}_{D}-E)=E^4-A E^2 +B, 
\end{equation}
 with 
 \begin{equation}
     A = (E^{ABS}_{1})^2+(E^{ABS}_{2})^2, \quad B= (E_{1}^{ABS}E_{2}^{ABS})^2.
 \end{equation}
A zero-energy WP requires 
\begin{equation}\label{eq: conditions for ZEWPs}
    B= 0,
\end{equation}
whereas a finite-energy WP requires 
\begin{equation}\label{eq: conditions for FEWPs}
A^2 - 4B = 0.
\end{equation}
The coefficients $A$ and $B$ can be expressed in terms of the microscopic parameters by matching powers of $E$ in Eq.~\eqref{eq: det}. The resulting formulas are lengthy, so below we give the normal-coupled double-QD limit and the purely SC-coupled limit which clarify why robust finite-energy WPs in our working model require both normal and SC interdot coupling.

\subsubsection{Normal-coupled double-QD model}\label{appendix: WPs in N-DQD model}
For the normal-coupled double-QD model, Eq.~\eqref{eq: conditions for ZEWPs} becomes
\begin{equation}\label{eq: DQD condition for zero energy WPs}
    (r_N^2 - \varepsilon_{1} \varepsilon_2 - |\gamma_1||\gamma_2|)^2 + (|\gamma_1| \varepsilon_2 - |\gamma_2| \varepsilon_1)^2 +4r_N^2 |\gamma_1||\gamma_2| \cos^2[(\psi_1 - \psi_2)/2]=0,
\end{equation}
The finite-energy crossing condition, Eq.~\eqref{eq: conditions for FEWPs}, becomes

\begin{equation}\label{eq: DQD condition for finite enegy WPs}
\left(|\gamma_1|^2-|\gamma_2|^2+\varepsilon_1^2-\varepsilon_2^2\right)^2
+4r_N^2\Bigl[\left(\varepsilon_1+\varepsilon_2\right)^2 +\left(|\gamma_1|\cos\psi_1-|\gamma_2|\cos\psi_2\right)^2+
\left(|\gamma_1|\sin\psi_1-|\gamma_2|\sin\psi_2\right)^2\Bigl] = 0.
\end{equation}
As shown in Ref.~\cite{Teshler2023}, Eq.~\eqref{eq: DQD condition for zero energy WPs} can be satisfied over a broad parameter range. The finite-energy crossing condition is much more restrictive and, in particular, requires $\varepsilon_1=-\varepsilon_2$. Thus, robust finite-energy WPs are excluded in the generic normal-coupled double-QD case considered in the main text, apart from such fine-tuned limits.

\subsubsection{Purely SC-coupled double-QD model}\label{appendix: WPs in SC-DQD model}
For the purely SC-coupled limit, the zero-energy condition is
\begin{equation}
\left(|r_{\mathrm{SC}}|^2+\varepsilon_1\varepsilon_2-|\gamma_1||\gamma_2|\right)^2
+\left(|\gamma_1|\varepsilon_2+|\gamma_2|\varepsilon_1\right)^2 +4|\gamma_1||\gamma_2| |r_{\mathrm{SC}}|^2
\sin^2\left(\frac{\psi_1+\psi_2-2\psi_{\mathrm{SC}}}{2}\right) =0.
\end{equation}
The finite-energy condition is
\begin{equation}\label{eq: SC DQD condition for finite energy WPs}
\begin{aligned}
&\left(
|\gamma_1|^2+\varepsilon_1^2-|\gamma_2|^2-\varepsilon_2^2
\right)^2 \\
&\quad +4|r_{\mathrm{SC}}|^2\Bigl[
\left(\varepsilon_1-\varepsilon_2\right)^2
+\left(
|\gamma_1|\cos\psi_1+|\gamma_2|\cos(2\psi_{\mathrm{SC}}-\psi_2)
\right)^2 \\
&\hspace{4.2cm}
+\left(
|\gamma_1|\sin\psi_1+|\gamma_2|\sin(2\psi_{\mathrm{SC}}-\psi_2)
\right)^2
\Bigr]=0.
\end{aligned}
\end{equation}
The finite-energy condition is again restrictive. For instance, it requires $\varepsilon_1 = \varepsilon_2$. As a result,  satisfying Eq.~\eqref{eq: SC DQD condition for finite energy WPs} implies fine-tuning of the system's parameters, whereas allowing both normal hopping and nonlocal SC coupling removes these fine-tuning constraints and produces finite-energy WPs over a broader parameter range, as used in the main text.

\section{Linear response theory}
This appendix gives the finite-temperature linear-response derivation used in Sec.~\ref{sec: linear response}.
\subsection{ABS susceptibility at finite temperature}\label{appendix: ABS susceptibility}

First, we derive Eq.~\eqref{eq: susceptibility} from the Kubo formula. A phase
variation enters the effective Hamiltonian as
\begin{equation}\label{eq: perturbation}
\delta \hat H^{\mathrm{eff}}_{D}(t)
=
\sum_\alpha
\frac{\partial \hat H^{\mathrm{eff}}_{D}}{\partial\varphi_\alpha}
\delta\varphi_\alpha(t)
=
\sum_\alpha
\frac{\hbar}{2e}\hat I_\alpha\,\delta\varphi_\alpha(t),
\end{equation}
where
\begin{equation}
\hat I_\alpha
=
\frac{2e}{\hbar}
\frac{\partial \hat H^{\mathrm{eff}}_{D}}{\partial\varphi_\alpha}
\end{equation}
is the current operator in lead \(\alpha\). In frequency space, the Josephson
relation gives
\begin{equation}
\delta\varphi_\alpha(\omega)
=
\frac{2e}{\hbar}
\frac{V_\alpha(\omega)}{-i\omega},
\end{equation}
and therefore
\begin{equation}
\delta \hat H^{\mathrm{eff}}_{D}(\omega)
=
\sum_\alpha
\hat I_\alpha
\frac{V_\alpha(\omega)}{-i\omega}.
\end{equation}
Thus the ABS susceptibility in Eq.~\eqref{eq: current} is the current response to
\(V_\alpha(\omega)/(-i\omega)\), or equivalently the phase-response kernel multiplied
by \(2e/\hbar\).

To derive Eq.\ \eqref{eq: susceptibility}, we use the Matsubara technique followed by analytical continuation to real frequencies. We first recall the general Kubo response for an arbitrary operator $\hat A$. If the perturbation is
\begin{equation}
\delta \hat H(t)=\hat B\,f(t),
\end{equation}
then the linear response of $\hat A$ is determined by the imaginary-time correlator
\begin{equation}
\chi^{K}_{AB}(\tau)
=
-
\left\langle
T_\tau \delta \hat A(\tau)\delta \hat B(0)
\right\rangle ,
\end{equation}
where
\begin{equation}
\delta \hat A=\hat A-\langle \hat A\rangle,
\qquad
\delta \hat B=\hat B-\langle \hat B\rangle .
\end{equation}
Here $\hat A$ is an arbitrary operator whose response is measured, while $\hat B$ is the operator coupled to the external perturbation.

For the ABS susceptibility we take
\begin{equation}
\hat A=\hat I_\alpha,
\qquad
\hat B=\hat I_\beta .
\end{equation}
Therefore the Kubo part of the ABS response is
\begin{equation}
\chi^{ABS}_{\alpha\beta}(\tau)
=
-
\left\langle
T_\tau \delta \hat I_\alpha(\tau)\delta \hat I_\beta(0)
\right\rangle .
\end{equation}
In Matsubara-frequency representation this gives, for $i\Omega_m\neq0$,
\begin{equation}
\chi^{ABS}_{\alpha\beta}(i\Omega_m)
=
\sum_{k\neq k'}
I^\alpha_{kk'}I^\beta_{k'k}
\frac{n_F(E_k)-n_F(E_{k'})}
{i\hbar\Omega_m+E_k-E_{k'}} .
\end{equation}

The usual analytical continuation from nonzero Matsubara frequencies,
\begin{equation}
i\Omega_m\to\omega+i0,
\end{equation}
then gives Eq.\ \eqref{eq: susceptibility} from the main text at $\omega\neq0$.

\subsection{Reproduction of the transconductance quantization at $T=0$}\label{appendix: transconductance quantization}

Finally, we verify that the finite-temperature transconductance reduces to the quantized zero-temperature transconductance. The first-order term in frequency is obtained by expanding Eq.\ \eqref{eq: susceptibility} with respect to $\omega$:
\begin{equation}
\chi^{(1)}_{\alpha\beta}
=
-i\hbar
\sum_{k\neq k'}
\frac{n_F(E_k)-n_F(E_{k'})}
{(E_k-E_{k'}+i0)^2}
I^\alpha_{kk'}I^\beta_{k'k}.
\end{equation}
Generally, it can be written in the form of Eq.\ \eqref{eq: chi 1 finite temperature} with the finite-temperature Berry curvature introduced by Eq.\ \eqref{eq: finite temperature BC}.

At zero temperature, $n_F(E_k)=\theta(-E_k)$, and therefore
\begin{equation}
B^{\alpha\beta}_0
=
2\,{\rm Im}
\sum_{\substack{E_k<0\\ E_{k'}>0}}
\frac{
\langle k|\partial_\alpha H^{\mathrm{eff}}_{D}|k'\rangle
\langle k'|\partial_\beta H^{\mathrm{eff}}_{D}|k\rangle
}
{(E_k-E_{k'})^2}
=
\sum_{E_k<0}B^{\alpha\beta}_k .
\end{equation}
Thus $B^{\alpha\beta}_0=B^{\alpha\beta}$, and the transconductance becomes quantized, as in Eq.\ \eqref{eq: transconductance quantization}.

\section{Finite-temperature Berry-curvature maps}\label{sec: Visualization of finite-temperature Berry Curvature}
This appendix shows representative finite-temperature Berry-curvature maps whose BZ integrals give the finite-temperature Chern number: the symmetric single-QD and normal-coupled double-QD models for ground-state topology, the antisymmetric SC-coupled double-QD for individual-ABS topology, and a topologically trivial normal-coupled double-QD close to its transition. These maps distinguish the two thermal mechanisms discussed in Sec.~\ref{sec: mechanisms}.

\subsection{Finite-temperature Berry curvature of symmetric single-QD and normal-coupled double-QD models}\label{appendix: Visualization of BC for SQD and N-DQD models}

\begin{figure}[t]
\includegraphics[width=0.33\columnwidth]{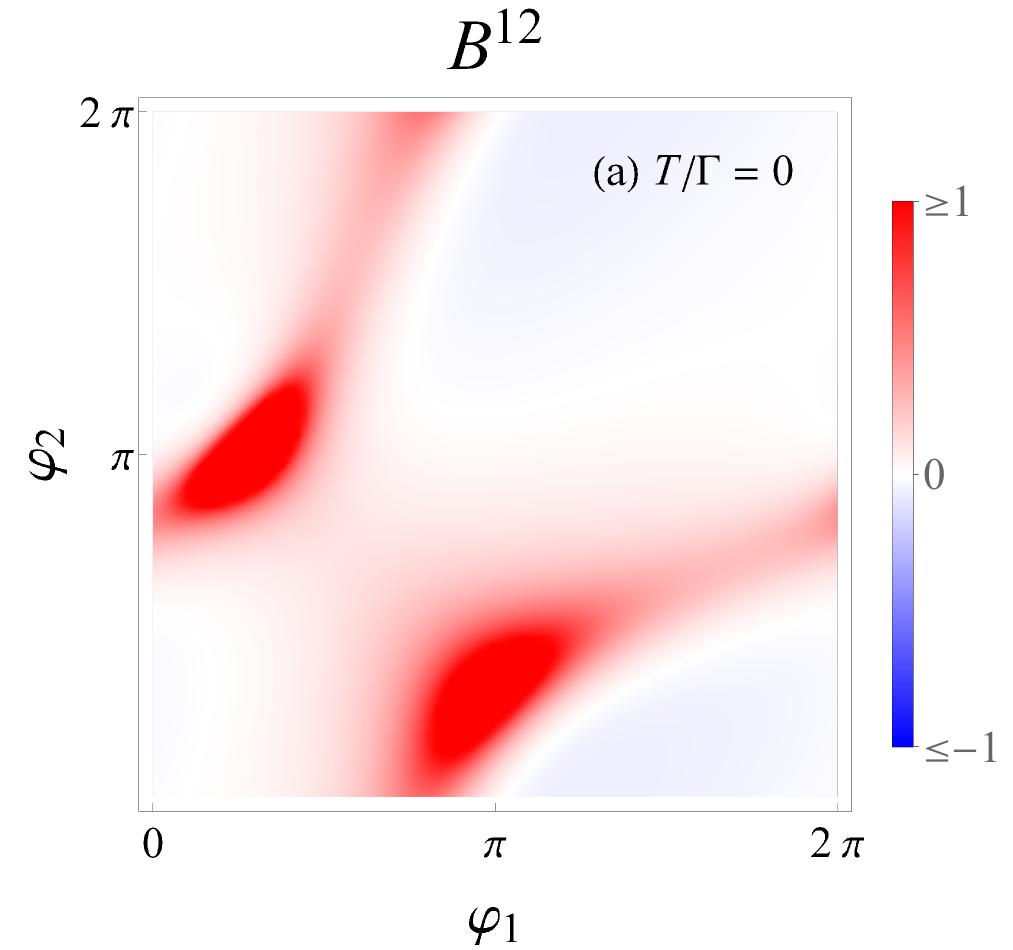}\centering \hfill 
\includegraphics[width=0.33\columnwidth]{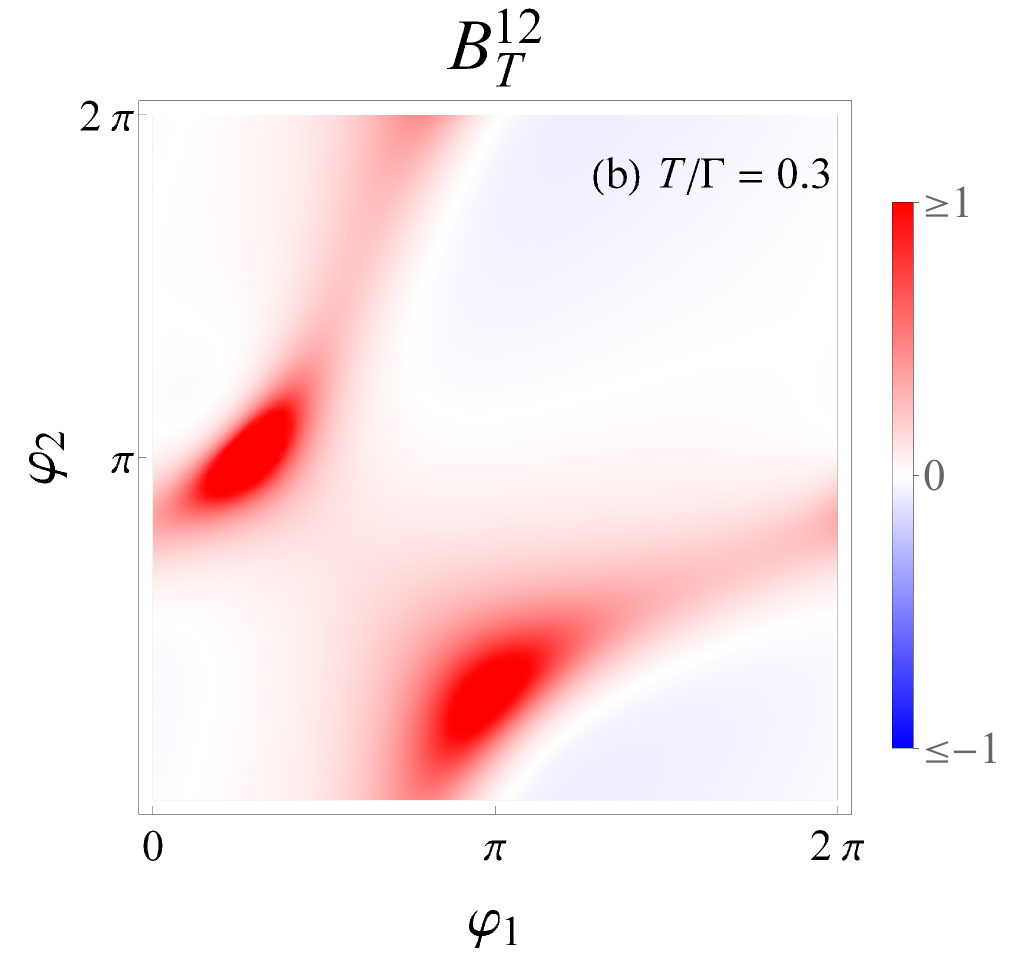}\centering \hfill 
\includegraphics[width=0.33\columnwidth]{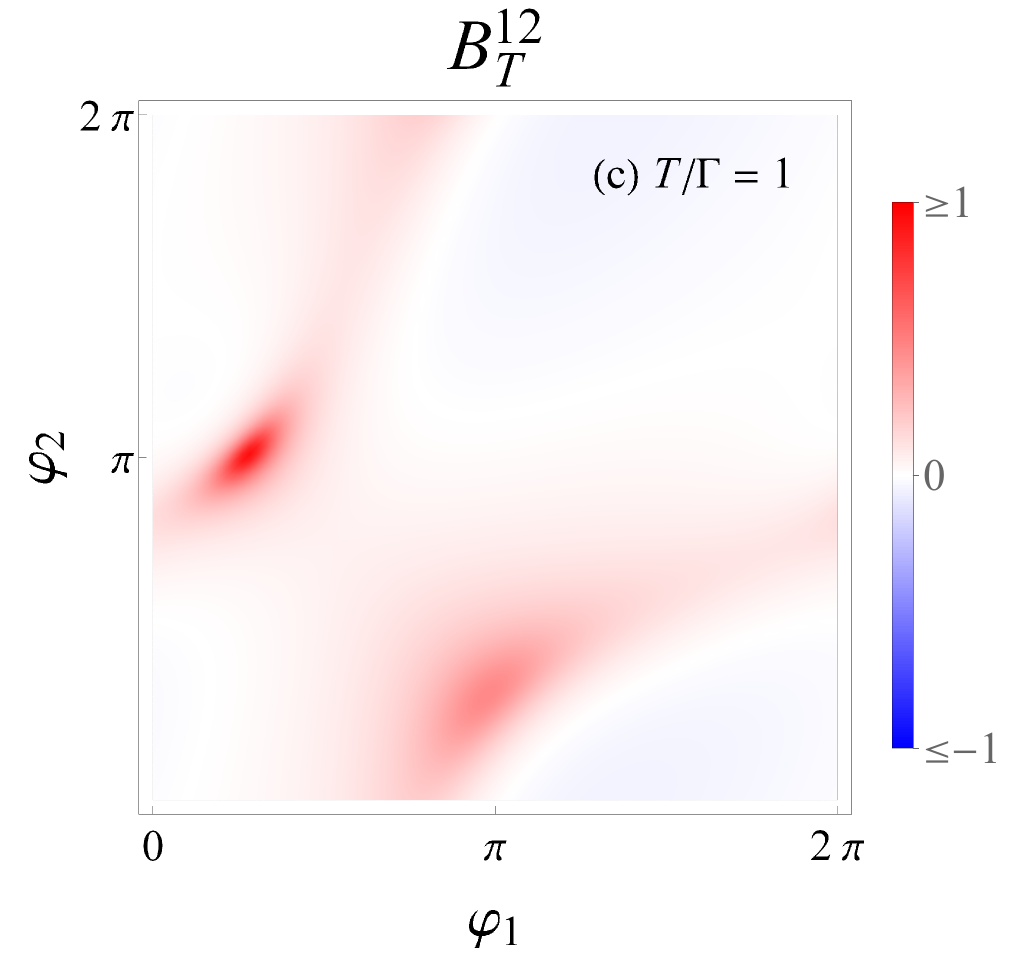}\centering \hfill \\
\includegraphics[width=0.33\columnwidth]{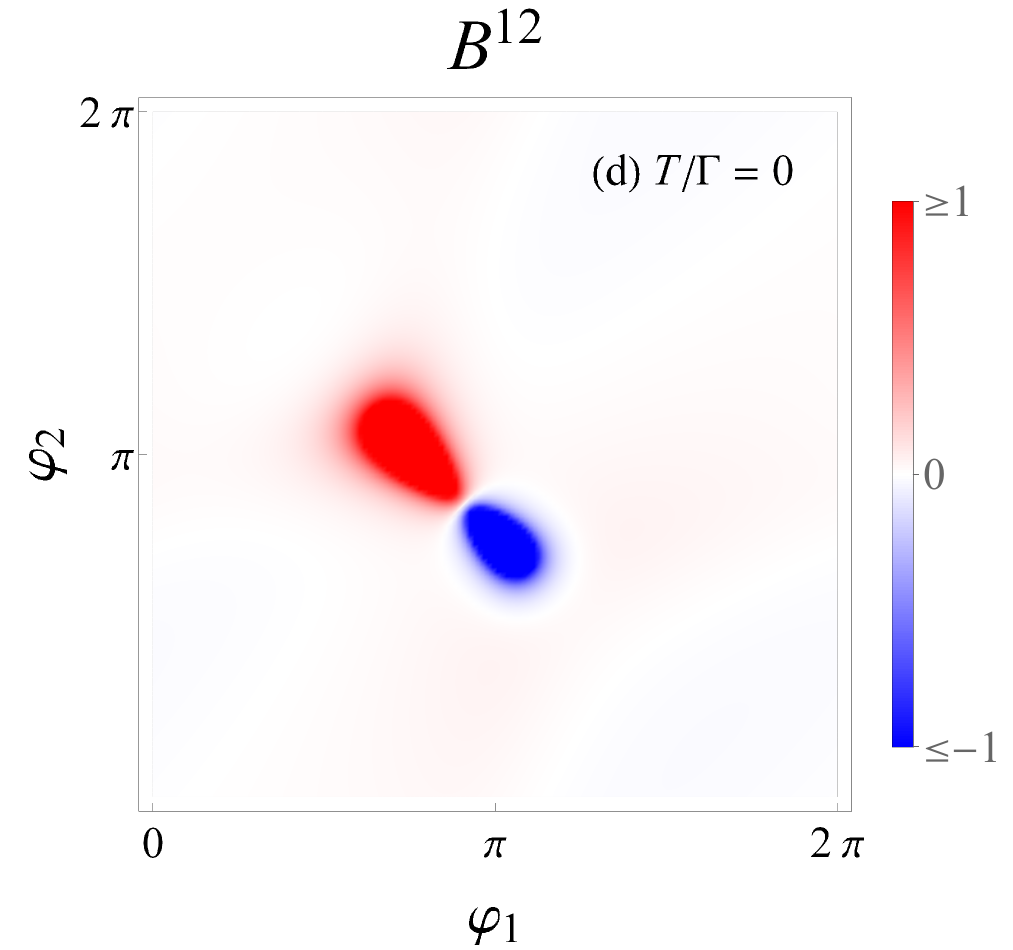}\centering \hfill 
\includegraphics[width=0.33\columnwidth]{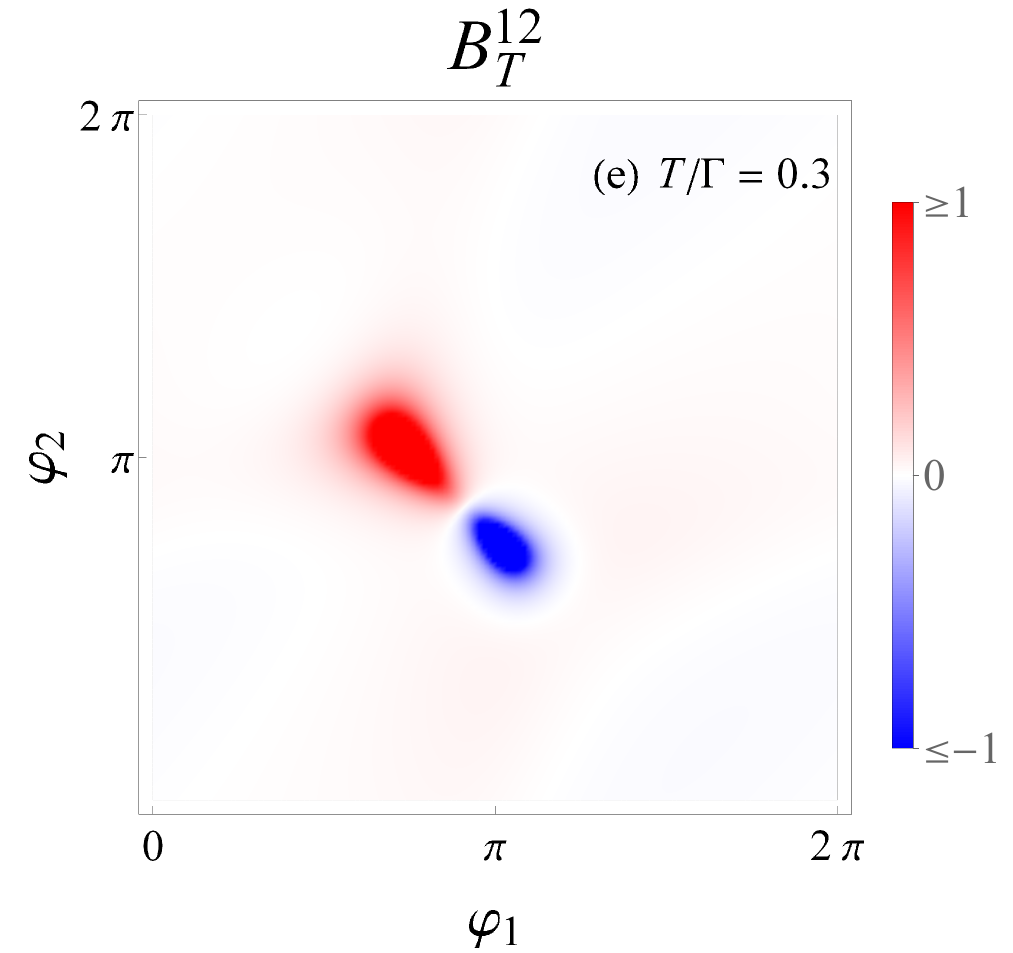}\centering \hfill 
\includegraphics[width=0.33\columnwidth]{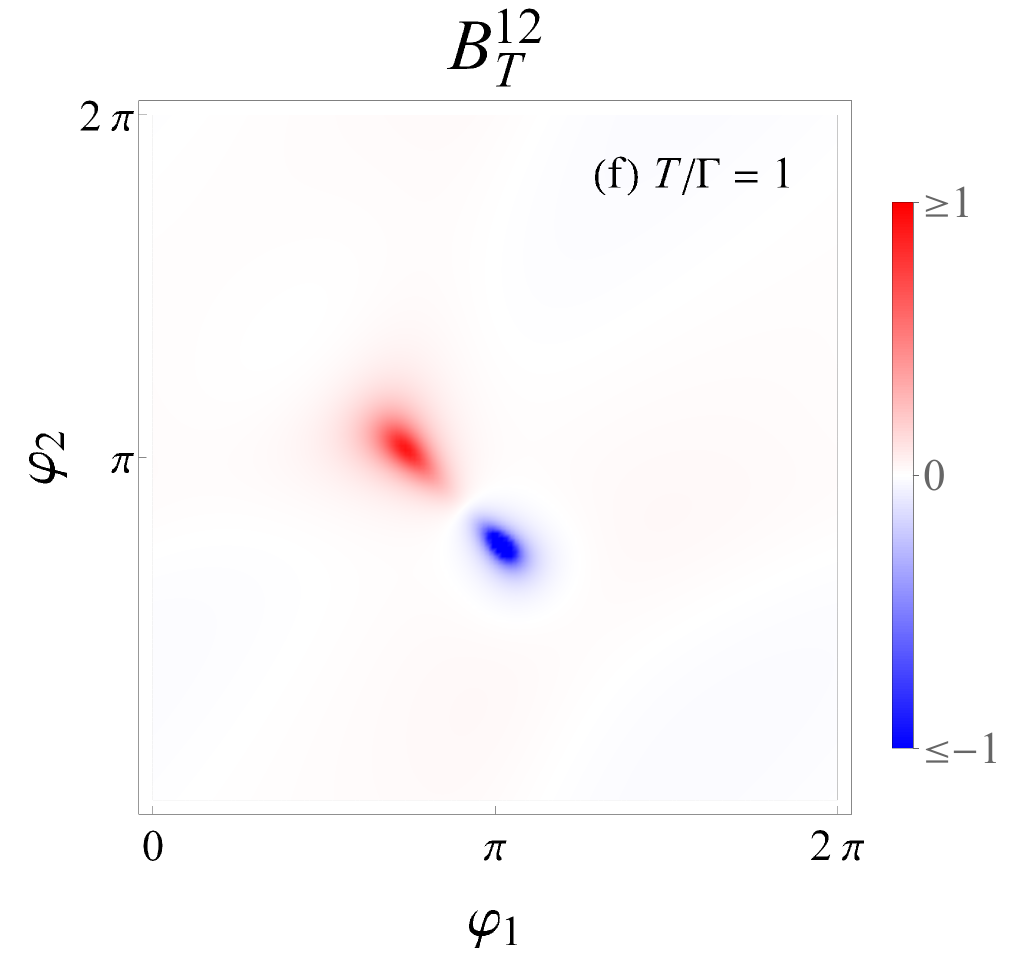}\centering \hfill 
\caption{Zero- and finite-temperature Berry-curvature maps for the symmetric single-QD model with $\varepsilon/\Gamma=-0.2$ and $\tau = 0.1$. Panels (a) and (d) show the zero-temperature Berry curvature $B^{12}$; panels (b), (c), (e), and (f) show $B_T^{12}$ at the temperatures indicated. The top row uses $\varphi_3=4$, inside the ground-state topological region [corresponding to Fig.~\ref{fig: finite temperature Chern number}(b)], and shows predominantly thermal suppression of the regions of positive Berry curvature. The bottom row uses $\varphi_3=5.5$, in the trivial region [Fig.~\ref{fig: finite temperature Chern number}(c)], where regions of positive and negative Berry curvature are suppressed at different rates.}
 \label{fig: FTBC in SQD model}
\end{figure}

\begin{figure}[t]
\centering
\includegraphics[width=0.33\columnwidth]{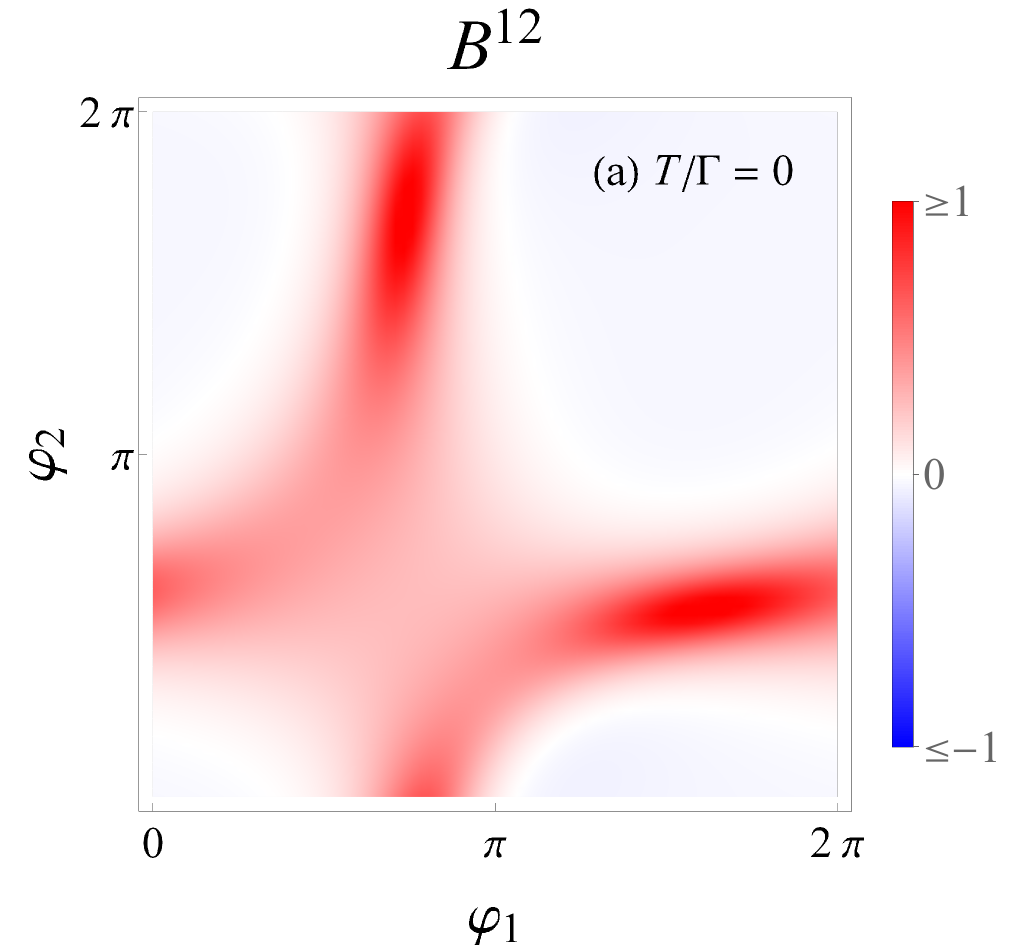}\centering \hfill 
\includegraphics[width=0.33\columnwidth]{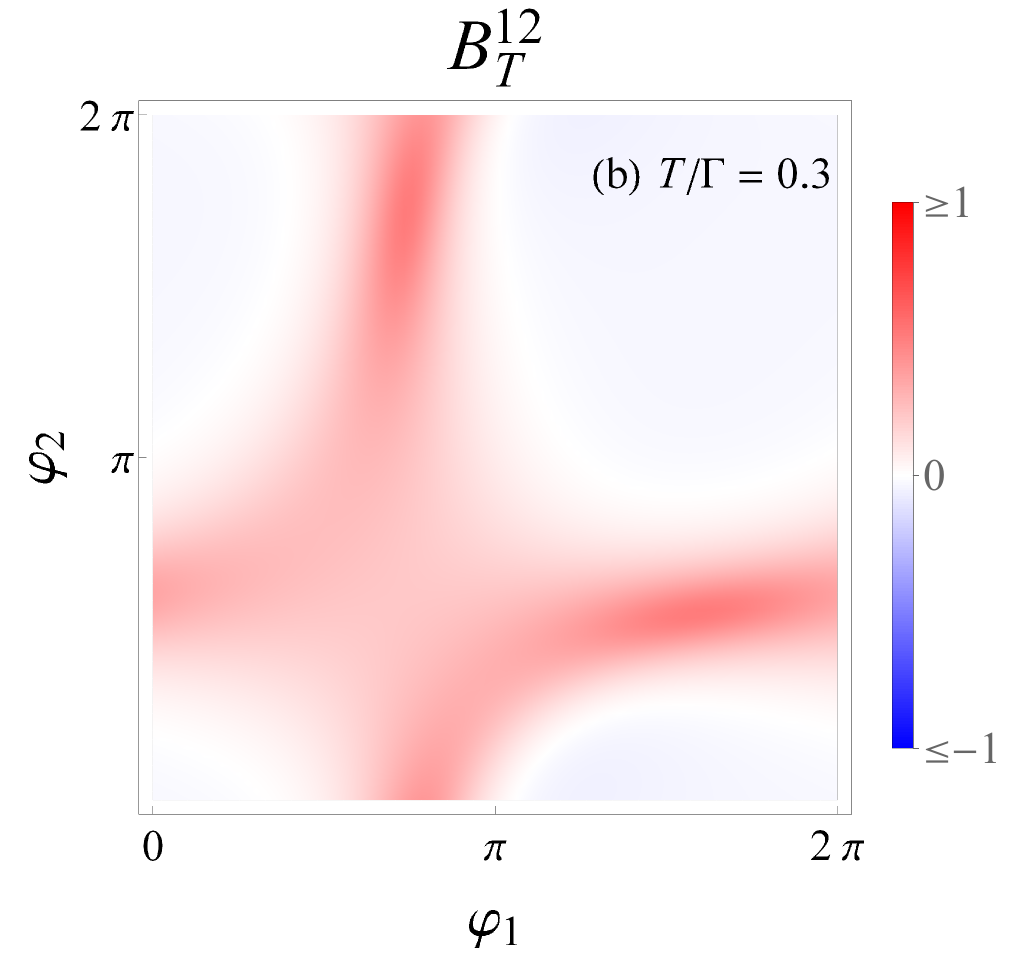}\centering \hfill 
\includegraphics[width=0.33\columnwidth]{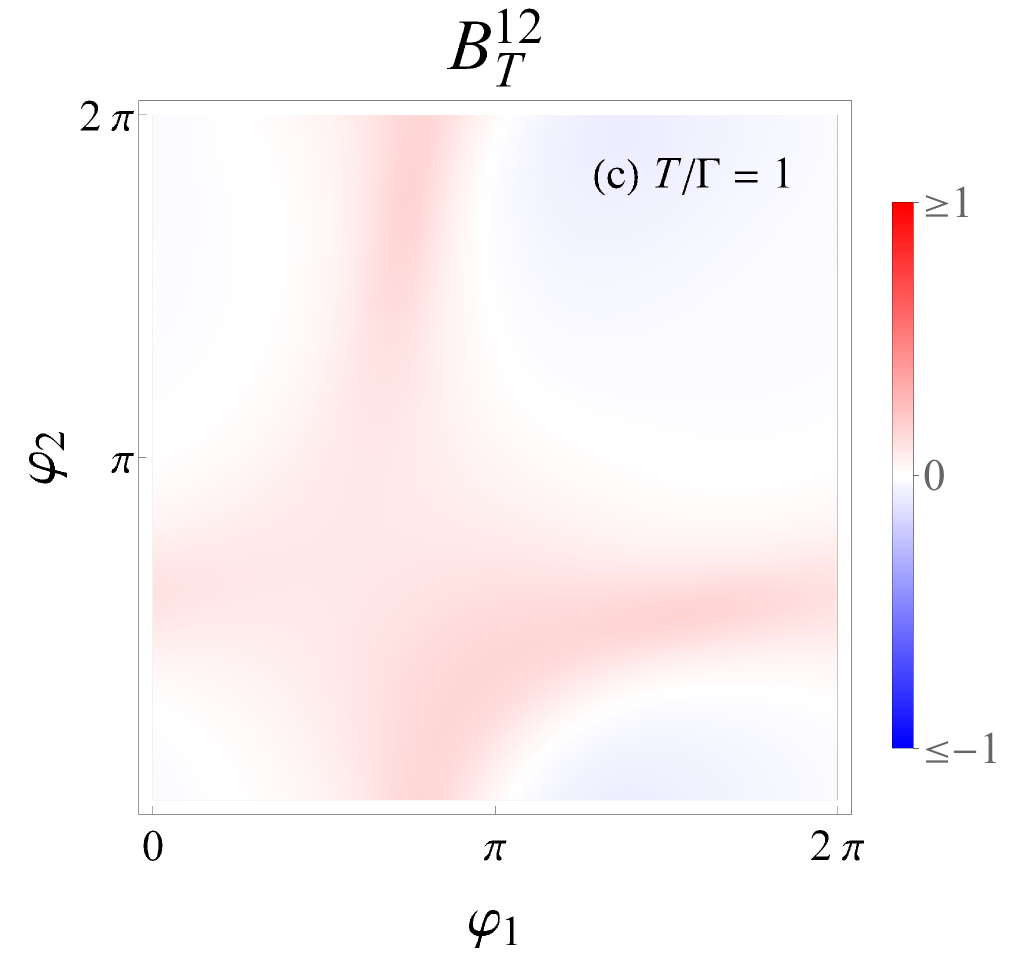}\centering \hfill \\
\includegraphics[width=0.33\columnwidth]{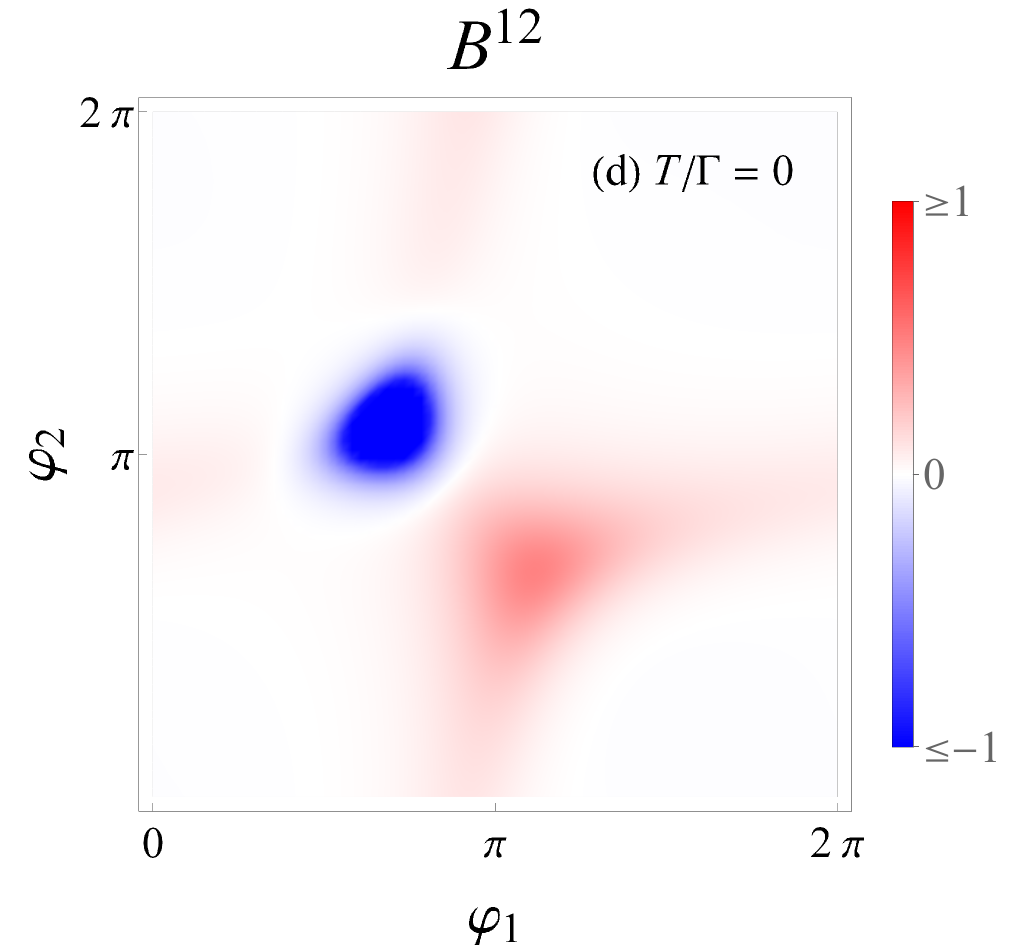}\centering \hfill 
\includegraphics[width=0.33\columnwidth]{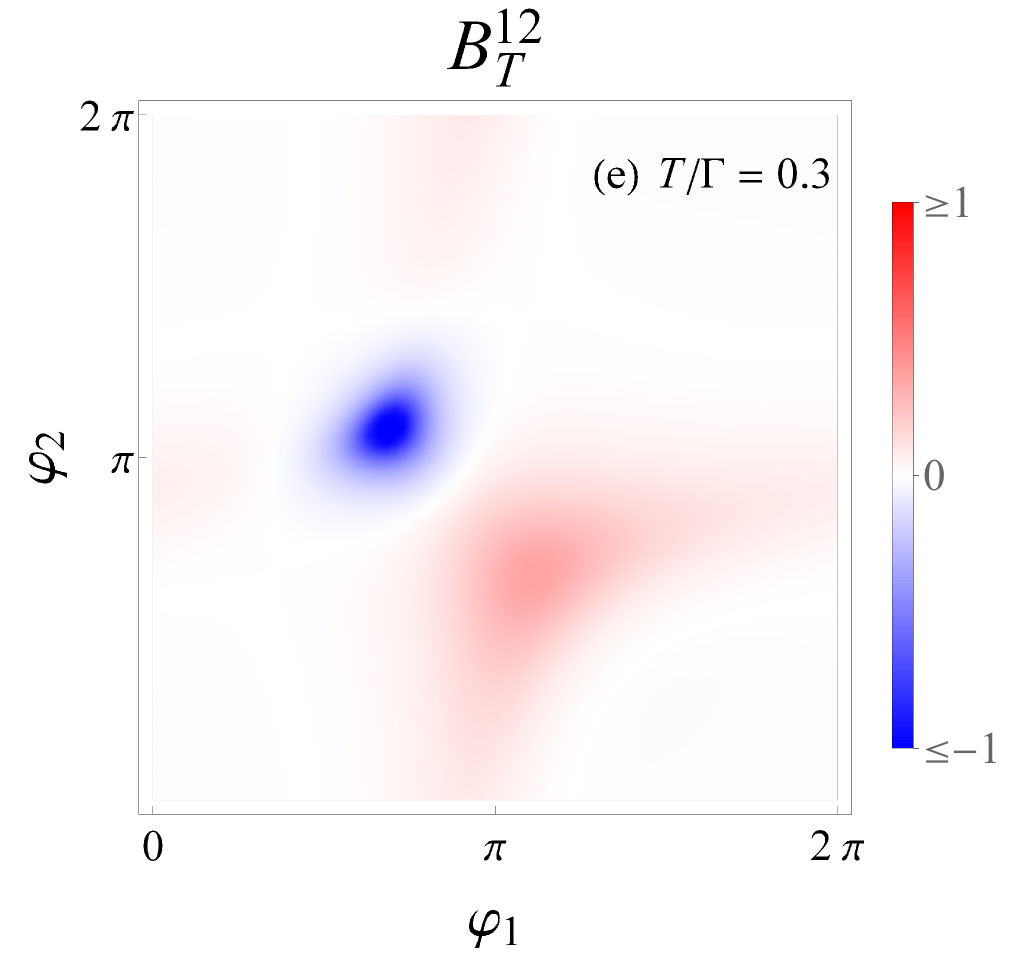}\centering \hfill 
\includegraphics[width=0.33\columnwidth]{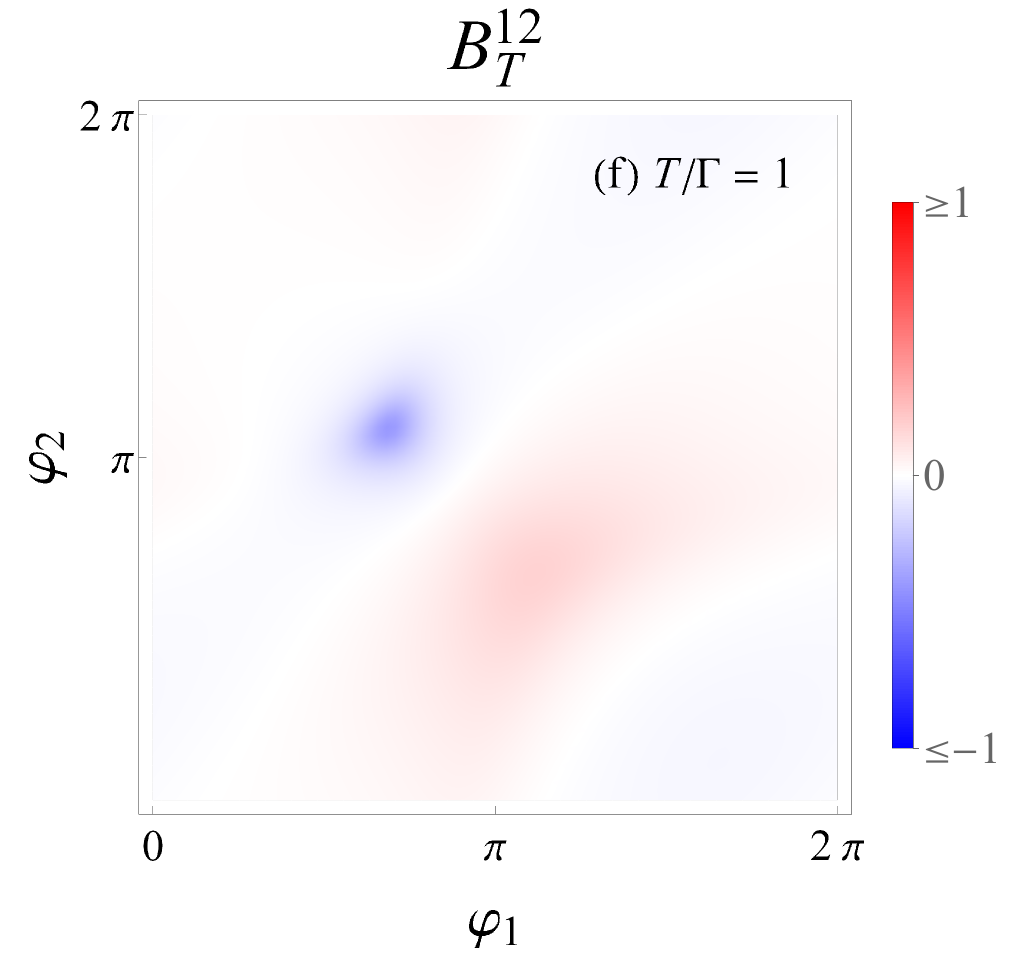}\centering \hfill
\caption{Zero- and finite-temperature Berry-curvature maps for the symmetric normal-coupled double-QD model with $\varepsilon/\Gamma=1$ and $r_N/\Gamma=1.5$. Panels (a) and (d) show $B^{12}$ at $T=0$; panels (b), (c), (e), and (f) show $B_T^{12}$ at the indicated temperatures. The top row uses $\varphi_3=4$, inside the ground-state topological region [Fig.~\ref{fig: finite temperature Chern number}(e)], where the dominant positive-Berry-curvature weight is thermally suppressed. The bottom row uses $\varphi_3=5.5$, in the trivial region [Fig.~\ref{fig: finite temperature Chern number}(f)], where negative and positive regions are suppressed at different rates.}
 \label{fig: FTBC in NC-DQD model}
\end{figure}

Figures~\ref{fig: FTBC in SQD model} and \ref{fig: FTBC in NC-DQD model} show the finite-temperature Berry curvature of the symmetric single-QD and normal-coupled double-QD models, respectively. As discussed in Sec.~\ref{sec: violation of quantization}, the thermal evolution differs between ground-state topological and trivial values of $\varphi_3$.

In the ground-state topological region [Figs.~\ref{fig: FTBC in SQD model}(a)--(c) and \ref{fig: FTBC in NC-DQD model}(a)--(c)], the Berry curvature integrates to $C^{12}=1$. Increasing temperature predominantly suppresses the regions of positive Berry curvature, while negative contributions change more weakly. The resulting loss of positive weight produces the monotonic decrease of $C_T^{12}$ shown in Figs.~\ref{fig: finite temperature Chern number}(b) and (e).

In the ground-state trivial region [Figs.~\ref{fig: FTBC in SQD model}(d)--(f) and \ref{fig: FTBC in NC-DQD model}(d)--(f)], positive- and negative-Berry-curvature contributions cancel at $T=0$, giving $C^{12}=0$. At low temperature, the negative contribution is suppressed more strongly on average, so $C_T^{12}$ initially increases. At higher temperature, the relative suppression changes and $C_T^{12}$ passes through a maximum before decreasing, as in Figs.~\ref{fig: finite temperature Chern number}(c) and (f).

 Although the single-QD and normal-coupled double-QD models show similar $C_T^{12}(\varphi_3)$ trends, their finite-temperature Berry-curvature maps differ. As discussed in Sec.~\ref{sec: SQD model first mechanism}, Eq.~\eqref{eq: FTBC in SQD model} makes the single-QD response purely suppressive. By contrast, the normal-coupled double-QD model can show phase-resolved enhancement [the appearance of the light-blue region in Fig.~\ref{fig: FTBC in NC-DQD model}(f), which is absent in Fig.~\ref{fig: FTBC in NC-DQD model}(d)] because several ABS contributions with different signs are thermally reweighted.
 
\subsection{Finite-temperature Berry curvature of an antisymmetric SC-coupled double-QD model}\label{appendix: Visualization of BC in antisymmetric SC-DQD model}

\begin{figure}[t]
\includegraphics[width=0.35\columnwidth]{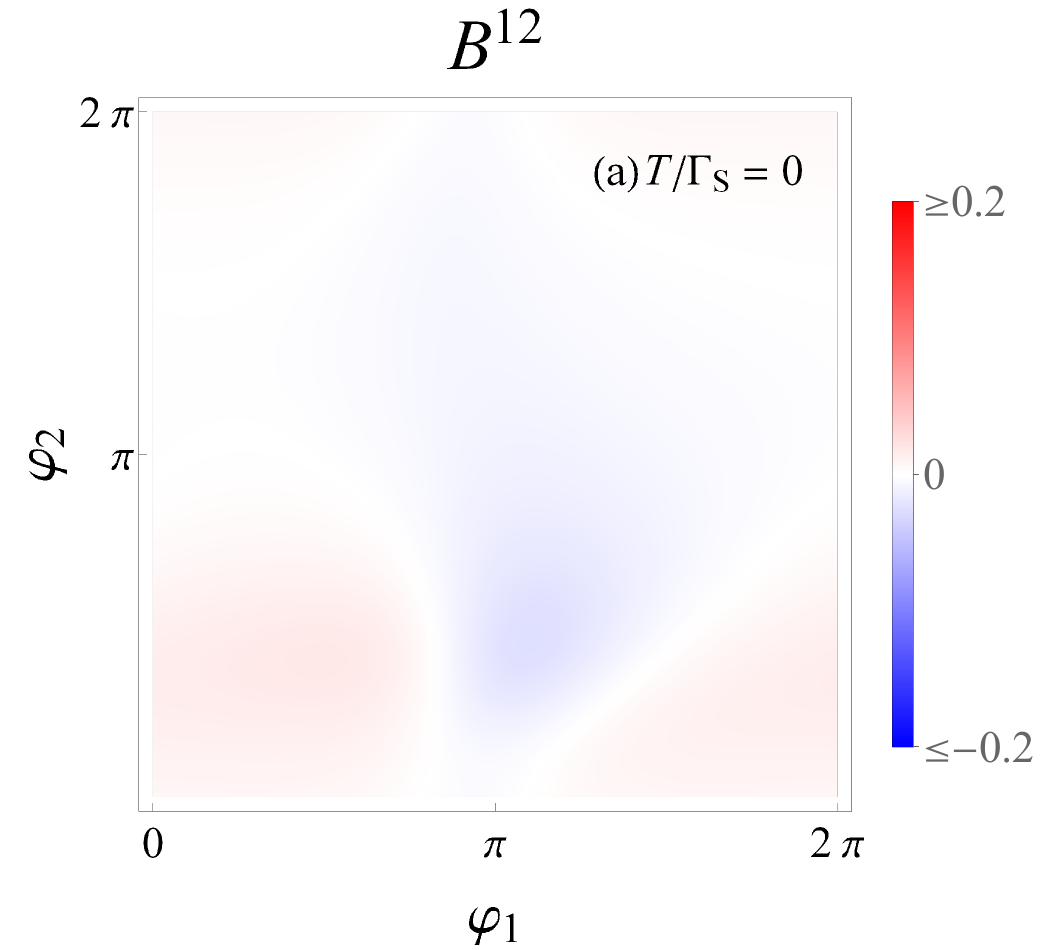}\centering \hspace{1 cm}
\includegraphics[width=0.35\columnwidth]{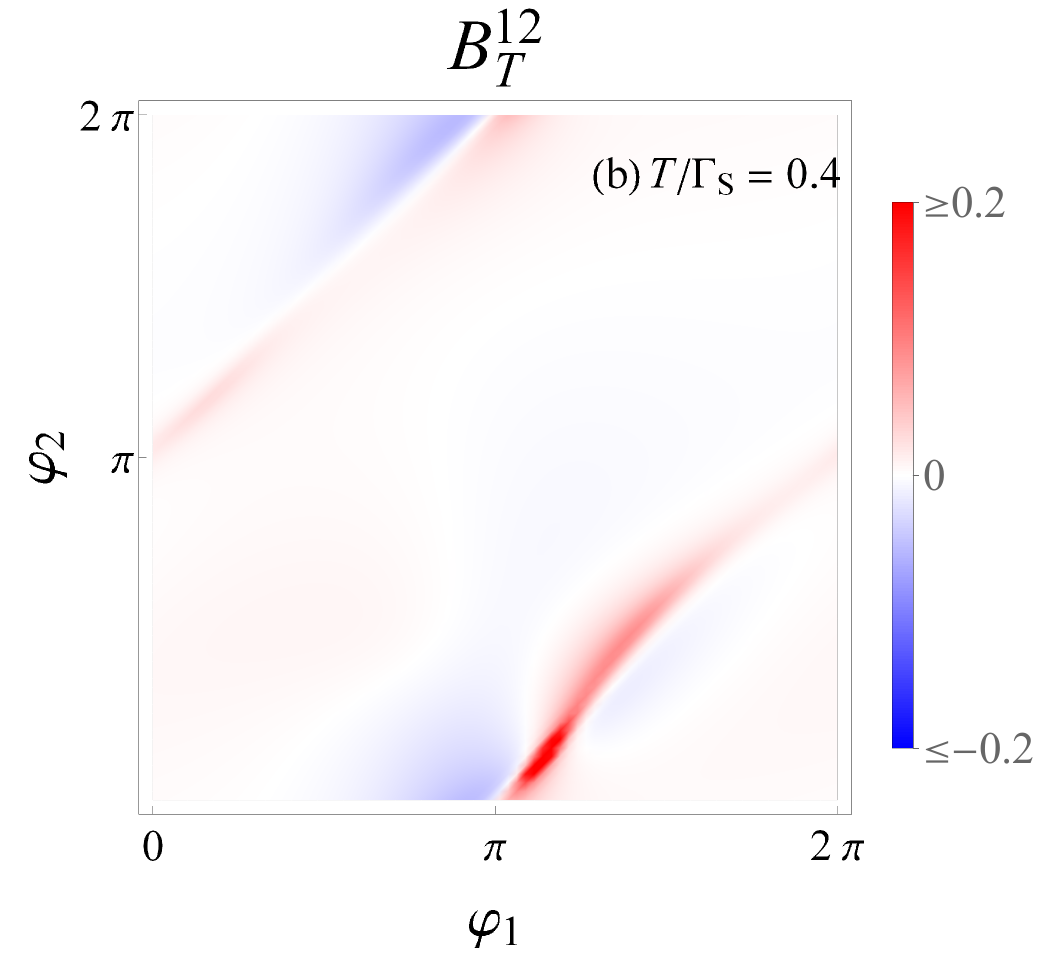}\centering \\
\includegraphics[width=0.35\columnwidth]{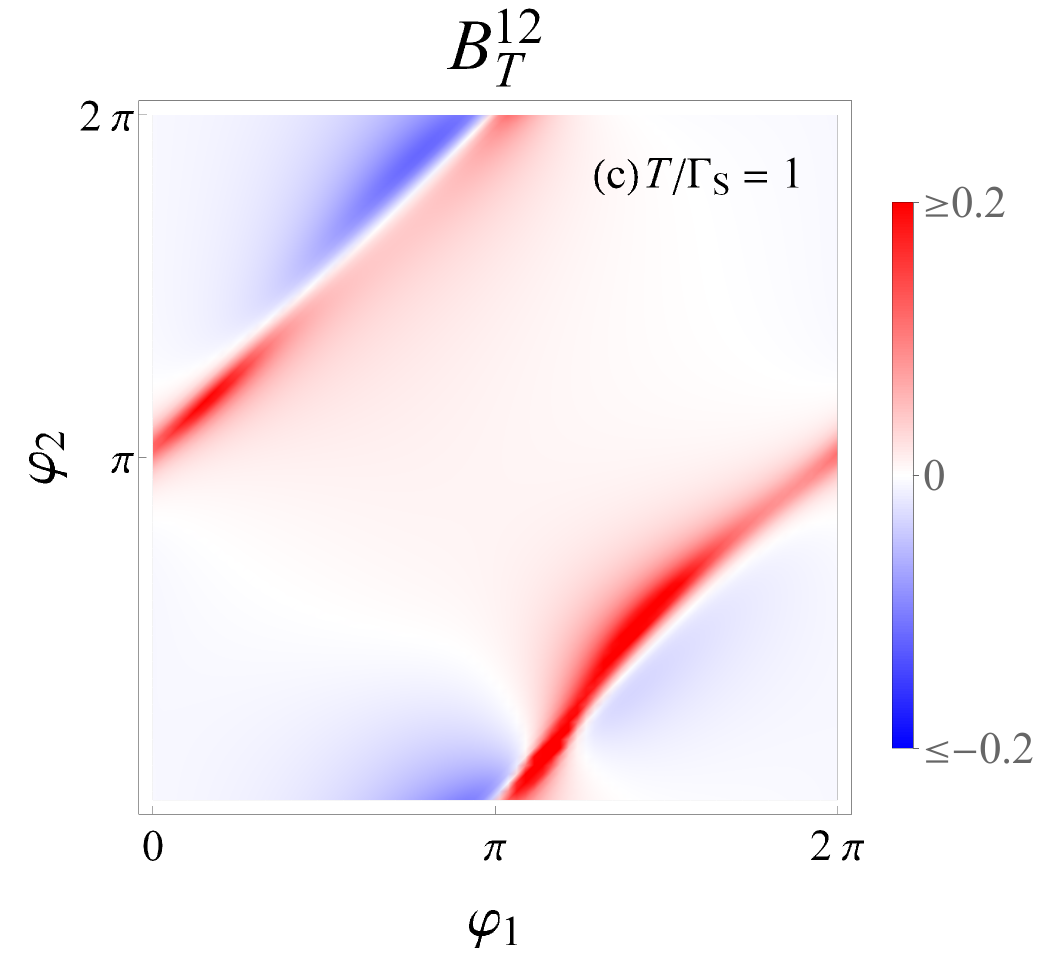}\centering \hspace{1 cm}
\includegraphics[width=0.35\columnwidth]{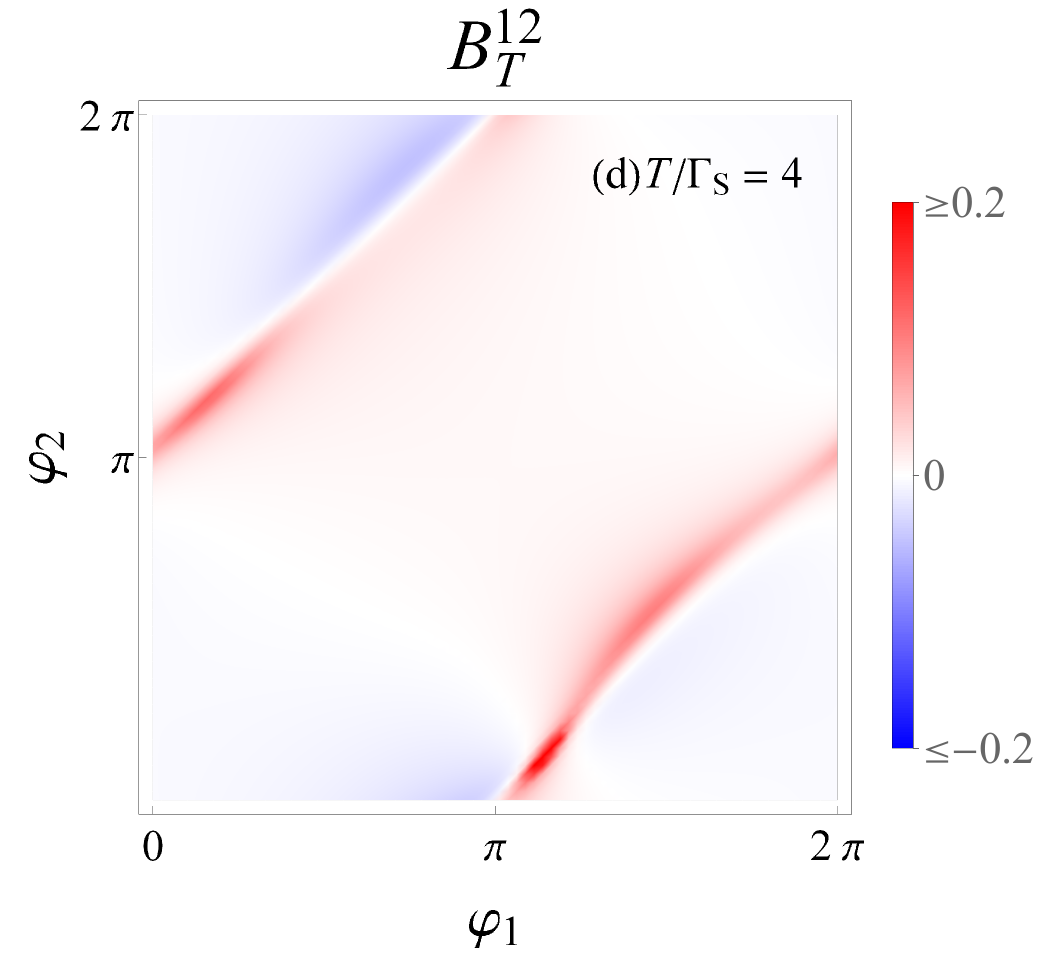}\centering
\caption{Zero- and finite-temperature Berry-curvature maps for the antisymmetric SC-coupled double-QD model with $\varepsilon/\Gamma_S=1.2$, $\Gamma_W/\Gamma_S=0.3$, $r_N/\Gamma_S=0.1$, and $\varphi_3=4$, corresponding to Fig.~\ref{fig: FTCN in the SC-DQD model}(b). Panel (a) shows $B^{12}$ at $T=0$; panels (b)--(d) show $B_T^{12}$ at increasing temperatures. The same-sector contribution produces the dominant temperature-induced change in the Berry curvature, which reaches a maximum and then decreases at higher temperature.}
 \label{fig: FTBC of SC-DQD}
\end{figure}

Figure~\ref{fig: FTBC of SC-DQD} shows the individual-ABS topological regime of the antisymmetric SC-coupled double-QD, corresponding to Fig.~\ref{fig: FTCN in the SC-DQD model}(b). At $T=0$, the Berry curvature is weak and integrates to $C^{12}=0$. Finite temperature generates finite-temperature Berry curvature through the same-sector contribution [Fig.~\ref{fig: FTBC of SC-DQD}(b)], which strengthens [Fig.~\ref{fig: FTBC of SC-DQD}(c)] and then weakens at higher $T$ [Fig.~\ref{fig: FTBC of SC-DQD}(d)]. This evolution produces the maximum and subsequent decay of $C_T^{12}(T)$ in Fig.~\ref{fig: FTCN in the SC-DQD model}(b).

\subsection{Finite-temperature Berry curvature of a topologically trivial normal-coupled double-QD model}\label{appendix: Visualization of FTBC in trivial model}

\begin{figure}[t]
\includegraphics[width=0.33\columnwidth]{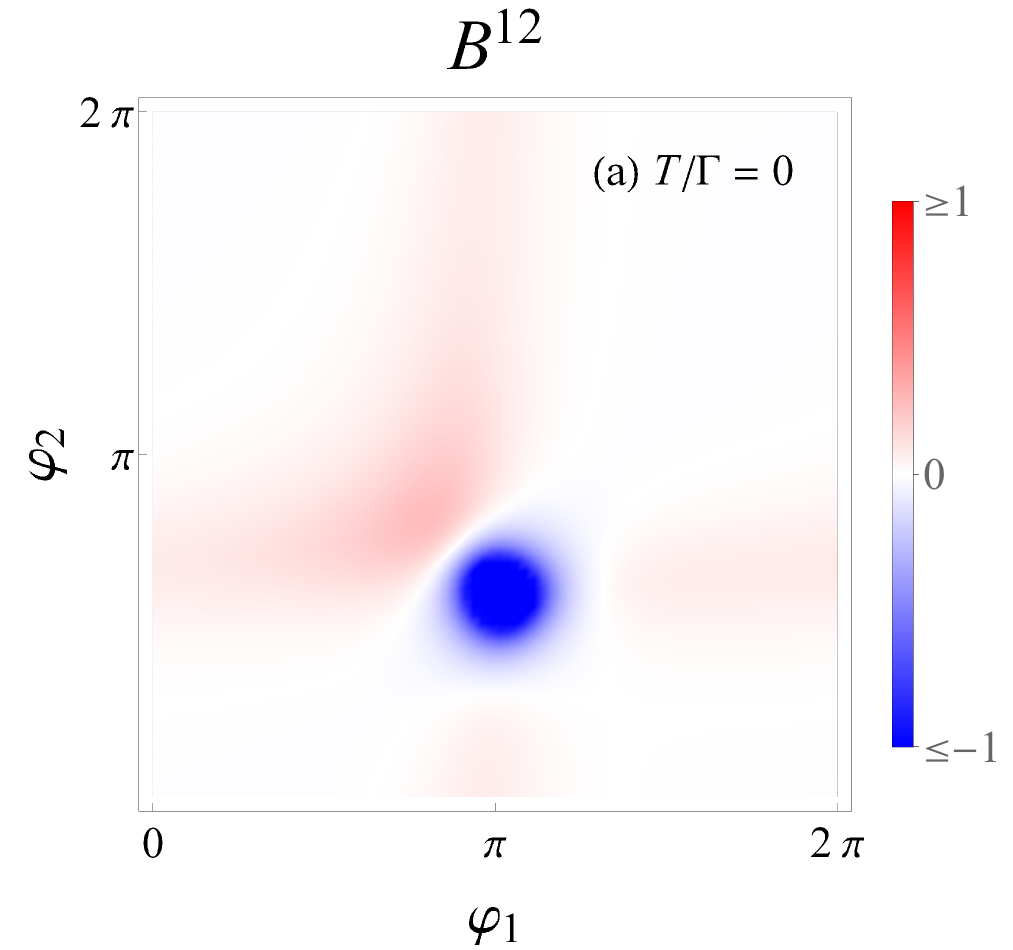}\centering \hfill 
\includegraphics[width=0.33\columnwidth]{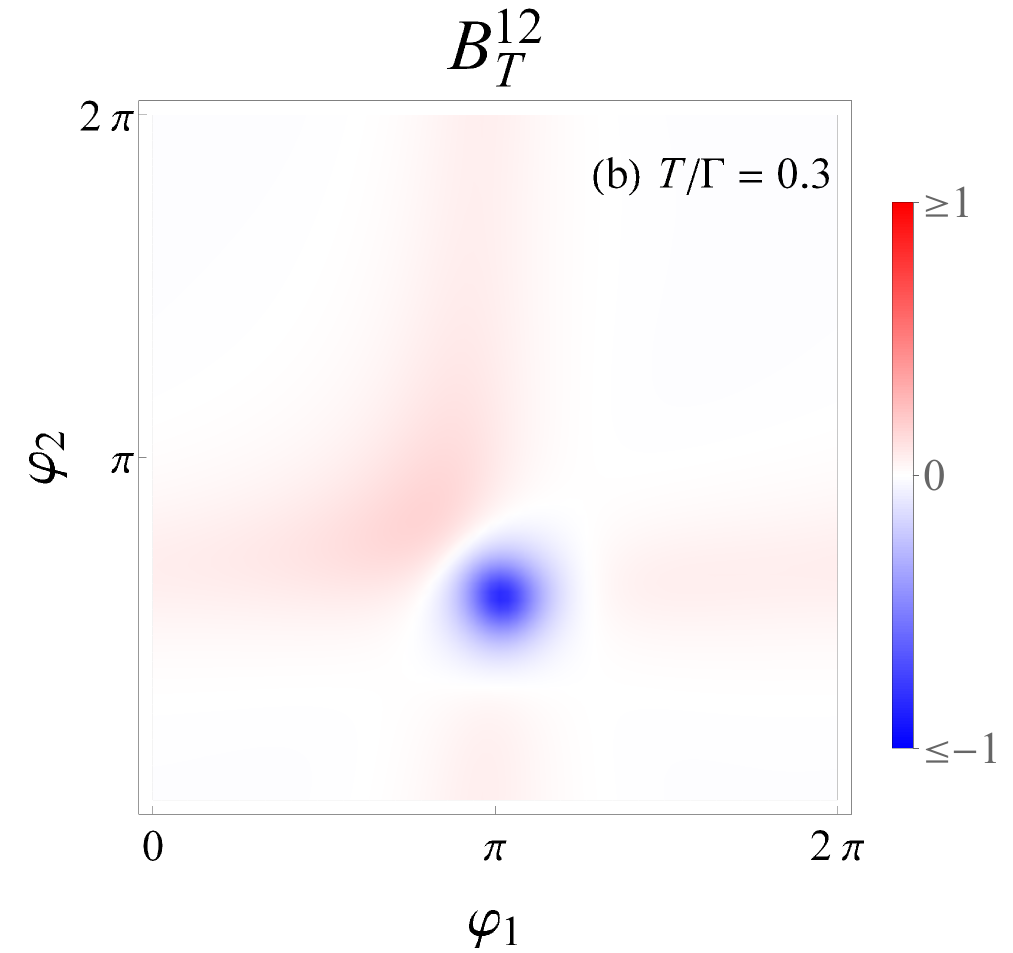}\centering \hfill 
\includegraphics[width=0.33\columnwidth]{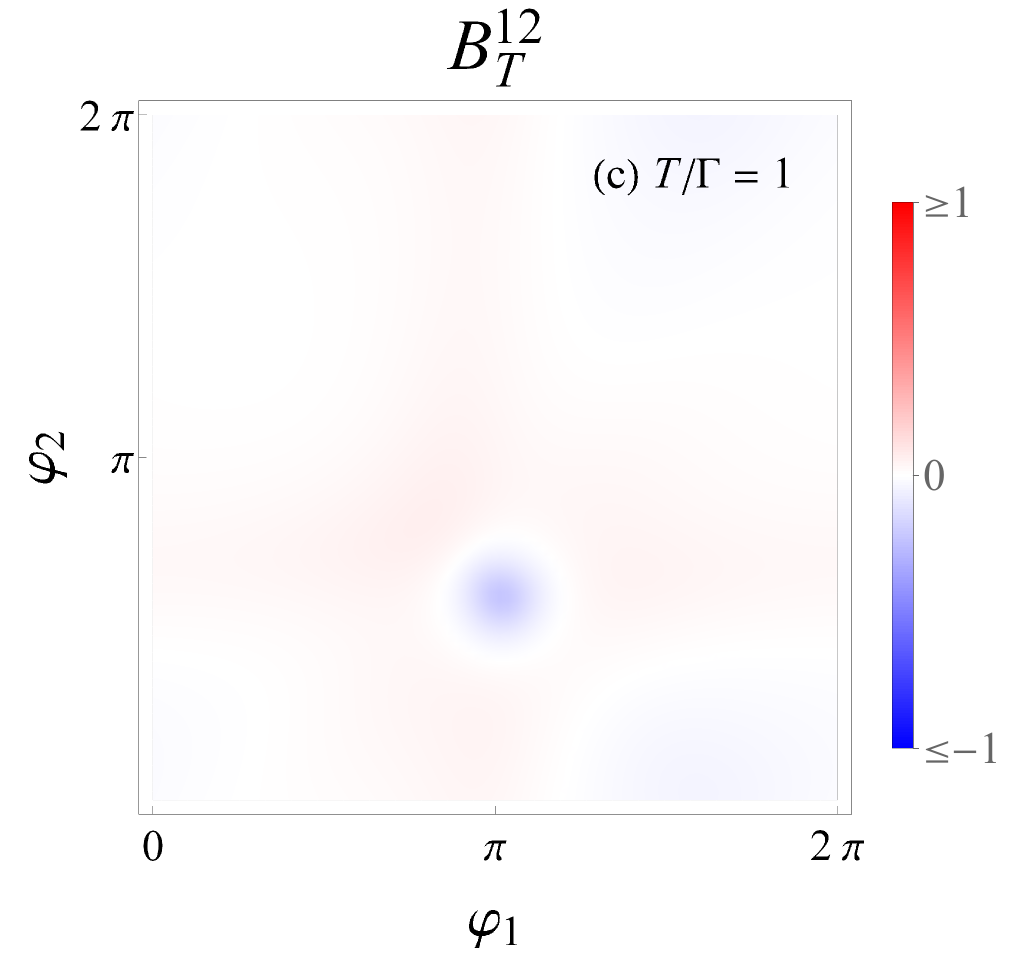}\centering 
\caption{Zero- and finite-temperature Berry-curvature maps for the symmetric normal-coupled double-QD model in the trivial regime. Panel (a) shows $B^{12}$ at $T=0$; panels (b) and (c) show $B_T^{12}$ at the indicated temperatures. The model parameters are $\varepsilon/\Gamma=1$, $r_N/\Gamma=0.8$, and $\varphi_3 = 5$, which corresponds to the black curve in Fig.~\ref{fig: FTCN in the SC-DQD model}(d). The dominant effect is unequal thermal reweighting of the regions of negative and positive Berry curvature.}
 \label{fig: FTBC in trivial N-DQD model}
\end{figure}

Finally, Fig.~\ref{fig: FTBC in trivial N-DQD model} shows the trivial normal-coupled double-QD corresponding to the black curve in Fig.~\ref{fig: FTCN in the SC-DQD model}(d). At $T=0$, regions of positive and negative Berry curvature cancel exactly, giving $C^{12}=0$. At low $T$, the negative contribution is suppressed more strongly, so $C_T^{12}$ initially increases; at higher $T$, the relative suppression changes and $C_T^{12}$ passes through a maximum before decreasing. Thus the dominant mechanism is unequal reweighting of pre-existing Berry curvature, rather than the thermally generated same-sector finite-temperature Berry curvature characteristic of the individual-ABS topological case.

\end{appendix}
\clearpage

\bibliography{Version_for_submission.bbl}

\end{document}